\documentclass{ws-mpla}
\usepackage[super]{cite}
\usepackage{graphicx}
\begin{document}

\markboth{PK Sahoo, Parbati Sahoo, Binaya K. Bishi, S. Ayg\"{u}n }
{Magnetized strange quark model with Big Rip singularity in $f(R,T)$ gravity}

\catchline{}{}{}{}{}

\title{Magnetized strange quark model with Big Rip singularity in $f(R,T)$ gravity}

\author{\footnotesize PK Sahoo}
\address{Department of Mathematics,\\ Birla Institute of
Technology and Science-Pilani, \\ Hyderabad Campus, Hyderabad-500078,
India\\
pksahoo@hyderabad.bits-pilani.ac.in}

\author{Parbati Sahoo}

\address{Department of Mathematics,\\ Birla Institute of
Technology and Science-Pilani, \\ Hyderabad Campus, Hyderabad-500078,
India\\
sahooparbati1990@gmail.com}

\author{Binaya K. Bishi}

\address{Department of Mathematics,\\ Visvesvaraya National Institute of Technology,\\ Nagpur-440010,India\\
binaybc@gmail.com}

\author{S. Ayg\"{u}n}

\address{Department of Physics,\\ \c{C}anakkale Onsekiz Mart University, Arts and Sciences Faculty,\\ Terzio\u{g}lu Campus, 17020, Turkey\\
saygun@comu.edu.tr}

\maketitle

\pub{Received (13 April 2017)}{Accepted (05 May 2017)}

\begin{abstract}
LRS (Locally Rotationally symmetric) Bianchi type-I magnetized strange quark matter cosmological model have been studied based on $f(R,T)$ gravity. The exact solutions of the field equations are derived with linearly time varying deceleration parameter which is consistent with observational data (from SNIa, BAO and CMB) of standard cosmology. It is observed that the model begins with big bang and ends with a Big Rip. The transition of deceleration parameter from decelerating phase to accelerating phase with respect to redshift obtained in our model fits with the recent observational data obtained by Farook et al. in 2017 \cite{Farooq17}. The well known Hubble parameter $H(z)$ and distance modulus $\mu(z)$ are discussed with redshift.

\end{abstract}

\ccode{PACS Nos.: 04.50.kd}

\section{Introduction}
Present universe constitutes various known and unknown components and matters. One of the most relevant matter called strange quark matter (SQM), which contains a large quantity of deconfined quark in $\beta$-equilibrium, with electric charge neutrality \cite{Alford07, BA sad07, Aktas11}. This deconfined quark matter is  composed of an equal number of up, down quarks and strange quarks. These quarks may be the true ground state of matter at high-density \cite{witten98}. Furthermore, it is believed that in the early stages of the universe, during phase transition of the universe, a transition called quark-hadron phase transition is occurred in which Quark Gluon Plasma (QGP) got transformed into hadron gas at temp $T\sim 200 MeV$. There are two approaches of strange quark matter creation: one is the aforesaid phase transition and another one is the strange matter made from the neutron star at ultra-high density \cite{bodmar71, Itoh70}. In the medium-dependent quark mass scale, the components of quark mass function are its chemical potential and temperature. These finite chemical potentials presently encounter serious problems. To resolve this, some effective phenomenological models are commonly used: MIT Bag model \cite{Chodos74} and the Nambu-Jona-Lasinio (NJL) model \cite{nambu, nambu61}. In bag model, all corrections of pressure and energy functions of SQM have been carried out through the introduction of ad-hoc bag function. In the framework of this model, Farhi and Jaffe \cite{frahi84} have studied the strange quark matter with its equation of state. In this model, the broken physical vacuums take place inside the hadrons on the basis of strong interaction theories. It gives essentially different vacuum energy densities inside and outside of hadron, and on the bag wall, the pressure of quarks are equilibrated through vacuum pressure and stabilizes the system. The equation of state (EoS) of the SQM depends upon the system pressure and various chemical potential components and the degenerate Fermi gasses. These Fermi gases are referred to as the replacement of quarks, which can survive only in a region with a vacuum energy density  $B_c$ (Bag constant). The unit of bag constant is $MeV/(fm)^3$ and it lies in the range 60-80 $MeV/(fm)^3$ \cite{Chakraborty2014}. In this study, we have assumed the value of $B_c$ to be 60 $MeV/(fm)^3$. By considering the quarks are massless and non-interacting in a simplified bag model, the quark pressure can be defined as  $P_q=\frac{\rho_q}{3},$  where  $\rho_q$ is the quark energy density.  Then the total energy density and pressure gave as:\\
\begin{eqnarray}
\rho_m=\rho_q+B_c \\
   P_m=P_q-B_c
\end{eqnarray}
Hence the EoS for strange quark matter \cite{Sotani2004} is
\begin{equation}
P_m=\frac{\rho-4B_c}{3}
\end{equation}
Again one of the most significant and widespread component of the universe is magnetic field, which is possessed by milk way galaxies and many other spiral galaxies along with some common properties of galaxy clusters \cite{Klein88, kronberg92, wolfe92, kronberg94, beck96, vallee04, carilli04}. Currently, the main focus of research is on the impact of strong magnetic field on the special properties of dense quark matter, neutron star matter \cite{Ducan92} and on the stability of strange quark matter \cite{fukusima13, kojo13, chakrabarty96}. Dey et al. \cite{dey98, dey99} have introduced the quark matter in strong magnetic field. When the electromagnetic scale becomes the order of the nuclear scales ($B\geq 10^{18}G$), then some substantial changes appear in the properties of strange matter. In ref \cite{chakrabarty96}, the quark matter has been studied in strong magnetic field with phenomenological bag model. It has been found that if the order of the strength of the magnetic field is greater than some critical value, the stability of SQM gets stronger. In literature, it has been accepted that the presence of magnetic field causes an anisotropy in pressure and the bag model can be considered as the best satisfactory approach for studying magnetized strange quark matter (MSQM) \cite{mart05,mart07}. Furthermore, the MSQM has been studied through the generalization of quasiparticle model \cite{Wen12}, in which they found a density and magnetic-field-dependent bag function which gives maximum saturated density at quantum chromodynamic (QCD) scale parameter greater than 123 MeV. \newline
Recent observations including supernovae-Ia, CMB, BAO, and LSS have predicted the conceptual evidence against the accelerated expansion of the universe \cite{Riess98, Perlmutter99, Riess99}. However, from the last century, the modern cosmology has been based only on the recent observations regarding accelerated expansion of the universe. The first reason behind this is the presence of an unknown form of matter and energy driven through negative pressure. Secondly, modification in the gravitational sector of the theory can also be considered as one of the good candidate for explaining the accelerated expansion of the universe. Some relevant alternative theories are Brans-Dicke (BD), scalar-tensor theories of gravitation, $f(R)$ gravity \cite{carroll04, nojiri07, bertolami07}, $f(T)$ gravity \cite{bengocheu09, linder10}, $f(G)$ gravity \cite{bamba10a, bamba10b, rodrigues14}, $f(R,G)$ gravity, where $R$, $T$ and $G$ are the scalar curvature, the torsion scalar and the Gauss-Bonnet scalar tensor theory respectively. The recent generalisation of $f(R)$ gravity along with trace of stress energy momentum tensor is known as $f(R,T)$ gravity which was proposed by Harko et al. in 2011 \cite{Harko2011}. In this theory, the matter lagrangian consists of an arbitrary function of the curvature scalar $R$ and the trace of the energy-momentum tensor $T$. In literature, many cosmological models have been studied in the framework of $f(R,T)$ gravity, either in presence of various matter distributions and metrics \cite{Moraes2015, Momeni15, Mubasher12, Myrzakulov12} or through energy conditions \cite{Sharif13}. Recently, many authors have also studied the properties of quark matter and strange quark matter in the account of general relativity as well as various modified gravitational theories. In general theory of relativity, the strange quark matter has been studied along with cosmic string cloud for axially symmetric space-time by Katore \cite{Katore15} and again attached to both string cloud and domain wall for spherically symmetric Kink space-time by Sahoo and Mishra \cite{Sahoo13a, Sahoo13b}. In $f(R)$ theory of gravity, the behaviour of quark and strange quark matter has been studied for Bianchi type-I and V space time by Yilmaz et al. \cite{Yilmaz12} and for Kantowski-Sachs metric by Adhav et al. \cite{Adhav15}. In particular, Agrawal and Pawar \cite{PKA17} have studied the quark and strange quark matter for a LRS Bianchi I metric in $f(R,T)$ gravity. They have solved the field equations using constant deceleration parameter. \newline
In the early seventies, cosmology was defined for searching two numbers: the present universe expansion rate $H_0$ and deceleration parameter (DP) $q_0$ \cite{Sandage70}. Among all the observations, the best fit and most natural model to specify the accelerated expansion of the universe is $\Lambda$CDM model. This model predicts the value of DP $q_0 \sim \frac{1}{2}$, while the other CDM
models predict $q_0 = 0$. Furthermore, the value of $q_0$ is determined to a precision of $\pm 0.2$ by the group of Supernovae
cosmology project and the high $z$ supernovae team in which the distant Type Ia supernovae ($z \sim 0.3-0.7$) is used as
standard candles. Since the fate of the universe remains still undefined, the reasonable accuracy of these values is limited to analyzing all. To predict the fate of the universe we need not only the current numeric values of the parameters but their time dependence is too. The Hubble parameter and DP are the simplest cosmographic parameters, which allow testing the coincidences between the cosmological model with cosmological principle. These can be defined through the Taylor series expansion of scale factor at the present time $t_0$.
\begin{equation}
a(t)=a(t_0)+\dot{a}(t_0)[t-t_0]+\frac{1}{2}\ddot{a}(t_0)[t-t_0]^2+.....
\end{equation}
It can be rewritten as
\begin{equation}
\frac{a(t)}{a(t_0)}=1+H_0 [t-t_0]+\frac{q_0}{2}H_{0}^2[t-t_0]^2+.....
\end{equation}
As we know that, the sign of the DP determines the difference between the actual age of the universe and Hubble time, decelerating or accelerating. Since we are interested in the expansion regime, then in the constant DP: if $q>0$, the
age of the universe will be less than the Hubble time and decelerate one, if $q=0$ expansion occurs at constant rate and the age equals to Hubble time, if $-1<q<0$ the acceleration of the universe exhibits power-law expansion, exponential if $q=-1$ and super-exponential if $q<-1$. In the context of constant DP, Berman \cite{berman83}, Berman and Gomide \cite{berman88} proposed the law of variation of Hubble parameter in general relativity that yields constant DP ($q=m-1, m\geq 0$ is a constant). The constant DP is commonly used by cosmologist in literature with various aspects. In order to make more detailed description of the kinematics of cosmological expansion, it is useful to consider various forms of time dependence deceleration parameter. One of the most popular form is known as linearly varying  deceleration parameter (LVDP). Linear parametrization of the DP represents quite naturally. The next logical step towards the behaviour of future model is either it  expands forever or ends with a Big Rip in finite future. This can be parametrized with redshift parameter $z$, cosmic scale factor $a$ and with cosmic time $t$. It is used in Bianchi type-V cosmological model with of holographic dark energy to escape the Big Rip singularity \cite{Sarkar14}. Singh et al. \cite{Singh13} have been studied the homogeneous and anisotropic Bianchi type-I cosmological model in the presence of viscous fluid source of matter, which starts with a big bang and ends in a Big Rip. The kinematical behaviour of LVDP along with null energy condition (NEC) has been explored in the framework of $f(R, T)$  gravity for Bianchi type -I and V space-time \cite{Zubair16}. Akarsu et al.  \cite{Akarsu14} have described the fate of the universe through parametrization $q = q_0+q_1(1- \frac{t}{t_0})$, which is linear in cosmic time $t$, along with two well-known  additional parametrization of the DP $q = q_0+q_1 z$ and $q = q_0+q_1(1-\frac{a}{a_0})$, where $z$ and $a$ are the redshift parameter and scale factor respectively. Furthermore, they have studied the dynamics of the universe in comparison with the standard $\Lambda$CDM model. \newline
The present work is an extension of the previous work of Sahoo and Sivakumar \cite{Sahoo2015} where the model was discussed for perfect fluid source. Here, we have considered the magnetic field coupled with strange quark matter and obtained the cosmological model for linearly cosmic time parametrization of the deceleration parameter. The article is is organized as follows: Section I contains the brief introduction and motivation regarding the present work. In Sec-II, we have derived the basic formalism of $f(R,T)$ gravity and the field equations. Thereafter, the solution of field equations are determined in sec-III by using LVDP. The discussion and graphical resolution of parameters are presented in sec-IV. Finally, in sec-V, the conclusion and perspective of our approach are outlined.
\section{Field equations in $f(R,T)$ gravity}
By considering the metric dependent Lagrangian density $L_m$, the respective field equation for $f(R,T)$ gravity are formulated from the Hilbert-Einstein variational principle in the following manner.
\begin{equation}
S=\int \sqrt{-g}\biggl(\frac{1}{16\pi G}f(R,T)+L_{m}\biggr)d^{4}x
\end{equation}%
where, $L_{m}$ is the usual matter Lagrangian density of matter source, $f(R,T)$ is an arbitrary function of
Ricci scalar $R$ and the trace $T$ of the energy-momentum tensor $T_{ij}$ of
the matter source, and $g$ is the determinant of the metric tensor $g_{ij}$. The energy-momentum tensor $T_{ij}$ from Lagrangian matter is defined in the form
\begin{equation}
T_{ij}=-\frac{2}{\sqrt{-g}}\frac{\delta (\sqrt{-g}L_{m})}{\delta g^{ij}}
\end{equation}%
and its trace is $T=g^{ij}T_{ij}$.\newline
Here, we have assumed that the matter Lagrangian $L_{m}$ depends only on the
metric tensor component $g_{ij}$ rather than its derivatives. Hence, we
obtain
\begin{equation}
T_{ij}=g_{ij}L_{m}-\frac{\partial L_{m}}{\partial g^{ij}}
\end{equation}%
By varying the action $S$ in Eq. (6) with respect to $g_{ij}$, the $f(R,T)$
gravity field equations are obtained as
\begin{equation}
f_{R}(R,T)R_{ij}-\frac{1}{2}f(R,T)g_{ij}+(g_{ij}\Box -\nabla _{i}\nabla
_{j})f_{R}(R,T)=8\pi T_{ij}-f_{T}(R,T)T_{ij}-f_{T}(R,T)\Theta _{ij}
\end{equation}%
where
\begin{equation}
\Theta _{ij}=-2T_{ij}+g_{ij}L_{m}-2g^{lm}\frac{\partial ^{2}L_{m}}{\partial
g^{ij}\partial g^{lm}}
\end{equation}
Here $f_{R}(R,T)=\frac{\partial f(R,T)}{\partial R}$, $f_{T}(R,T)=\frac{%
\partial f(R,T)}{\partial T}$, $\Box \equiv \nabla ^{i}\nabla _{i}$ where $%
\nabla _{i}$ is the covariant derivative.\newline
Contracting Eq. (9), we get
\begin{equation}
f_{R}(R,T)R+3\Box f_{R}(R,T)-2f(R,T)=(8\pi -f_{T}(R,T))T-f_{T}(R,T)\Theta
\end{equation}%
where $\Theta =g^{ij}\Theta _{ij}$.\newline
From Eqs (9) and (11), the $f(R,T)$ gravity field equations takes the form
\begin{multline}
f_{R}(R,T)\biggl(R_{ij}-\frac{1}{3}Rg_{ij}\biggr)+\frac{1}{6}f(R,T)g_{ij}= \\
8\pi -f_{T}(R,T)\biggl(T_{ij}-\frac{1}{3}Tg_{ij}\biggr)-f_{T}(R,T)\biggl(%
\Theta _{ij}-\frac{1}{3}\Theta g_{ij}\biggr)+\nabla _{i}\nabla _{j}f_{R}(R,T)
\end{multline}%
It must be mentioned here that the physical nature of the matter field through $%
\Theta _{ij}$ is used to form the field equations of $f(R,T)$ gravity. To construct different kinds of cosmological models according to the choice of matter source, Harko et al. \cite{Harko2011} constructed
three different frames of $f(R,T)$ gravity as
\begin{itemize}
\item $f(R,T)=R+2f(T)$
\item $f(R,T)=f_1(R)+f_2(T)$
\item $f(R,T)=f_1(R)+f_2(R)f_3(T)$
\end{itemize}
There are individual sets of field equations for each frames of $f(R,T)$ gravity. Here, we consider the first frame i.e $f(R,T)=R+2f(T)$ and the corresponding field equation is given as
\begin{equation}
R_{ij}-\frac{1}{2}Rg_{ij}=8\pi T_{ij}-2f'(T)T_{ij}-2f'(T)\Theta_{ij}+f(T)g_{ij}
\end{equation}
We consider the spatially homogeneous LRS Bianchi type-I metric as
\begin{equation}
ds^{2}=dt^{2}-A^{2}dx^{2}-B^2(dy^{2}+dz^{2})
\end{equation}
where $A, B$ are functions of cosmic time $t$ only.\\
The energy momentum tensor for magnetized strange quark matters is considered as \cite{CGT1997, JDB2007}
\begin{equation}
T_{ij}=(\rho+p+h^2)u_{i}u_{j}+\bigg(\frac{h^2}{2}-p \bigg)g_{ij}-h_i h_j
\end{equation}
where $u^i=(0,0,0,1)$ is the four velocity vector in co-moving coordinate system satisfying $u_iu_j=1$
and the magnetic flux $h^2$ is chosen in the $x$-direction due to $h_iu^i=0$. Here, $p$ is the proper pressure and $\rho$ is the energy density.\\
The field equation (13) with cosmological constant $\Lambda$ and $f(T)=\mu T$ can be written as
\begin{equation}
G_{ij}=[8\pi+2\mu]T_{ij}+[\mu \rho-\mu p+2\mu h^2+\Lambda]g_{ij}
\end{equation}
where $\mu$ is an arbitrary constant.\\
The set of field equations for the metric (14) are obtained as
\begin{eqnarray}
2H_1H_2+H_2^2=-(12\pi+5\mu)h^2-(8\pi+3\mu)\rho+\mu p-\Lambda\\
2\dot{H_2}+3H_2^2= (4\pi-\mu)h^2+(8\pi+3\mu)p-\mu \rho-\Lambda\\
\dot{H_1}+\dot{H_2}+H_1^2+H_2^2+H_1H_2=-(4\pi+3\mu)h^2+(8\pi+3\mu)p-\mu \rho-\Lambda
\end{eqnarray}
$H_1=\frac{\dot{A}}{A}$ and $H_2=\frac{\dot{B}}{B}$ are the directional Hubble parameters with $H=\frac{H_1+2H_2}{3}$ is the mean Hubble parameter. The dot represent derivatives with respect to time $t$.\\
For the metric (14), the scalar expansion $\theta$ and shear scalar $\sigma$ are defined as
\begin{equation}
\theta=3H=H_1+2H_2
\end{equation}
\begin{equation}
\sigma^2=\frac{1}{3}(H_1-H_2)^2
\end{equation}
\section{Solutions of Field equations}
The set of field equations (17)-(19) have $A, B, \rho, p, h^2,$ and $\Lambda$ having six unknowns with three equations. In order to get physically viable models of the universe which are consistent with the observations, we have considered the following assumptions:
\begin{enumerate}
\item Initially, the linear relationship between the directional Hubble parameters $H_1$ and $H_2$ as
\begin{equation}
H_1=nH_2
\end{equation}
where $n\geq 0$ is an arbitrary constant which takes care about the anisotropy nature of the model. The above equation yields the shear scalar $\sigma$ is proportional to the scalar expansion $\theta.$
\item Secondly, the equation of state (EoS) for strange quark matter as
\begin{equation}
p=\frac{\rho-4B_c}{3}
\end{equation}
where $B_c$ is bag constant \cite{Sotani2004}.
\item Finally, the LVDP $q$.
\end{enumerate}
Using the first assumption the field equation (17)-(19) take the form
\begin{eqnarray}
\frac{9(2n+1)}{(n+2)^2}H^2=-(12\pi+5\mu)h^2-(8\pi+3\mu)\rho+\mu p-\Lambda\\
\biggl[\frac{27}{(n+2)^2}-\frac{6(1+q)}{n+2}\biggr]H^2= (4\pi-\mu)h^2+(8\pi+3\mu)p-\mu \rho-\Lambda\\
\biggl[\frac{9(n^2+n+1)}{(n+2)^2}-\frac{3(n+1)(1+q)}{n+2}\biggr]H^2=-(4\pi+3\mu)h^2+(8\pi+3\mu)p-\mu \rho-\Lambda
\end{eqnarray}
After using the EoS from equation (23), we obtain the following values
\begin{equation}
h^2=\frac{3(n-1)(q-2)}{2(4\pi+\mu)(n+2)}H^2
\end{equation}
\begin{equation}
\rho=\frac{-3}{4(4\pi+\mu)}\biggl[\frac{9(n-1)}{(n+2)^2}+\frac{3(3+qn-2n)}{(n+2)}\biggr]H^2+B_c
\end{equation}
\begin{equation}
p=\frac{-1}{4(4\pi+\mu)}\biggl[\frac{9(n-1)}{(n+2)^2}+\frac{3(3+qn-2n)}{(n+2)}\biggr]H^2-B_c
\end{equation}
\begin{multline}
\Lambda=\biggl[\frac{3[(12n\pi+3n\mu-n^2 \mu+24\pi+10\mu)q]}{2(4\pi+\mu)(n+2)^2}+\frac{(-26\mu+18n\mu+6n^2 \mu-76\pi)}{2(4\pi+\mu)(n+2)^2}\biggr]H^2\\-(8\pi+4\mu)B_c
\end{multline}
Here, we have considered the linearly time varying deceleration parameter in the form \cite{Akarsu2012}
\begin{equation}
q(t)=-\frac{a\ddot{a}}{\dot{a}^2}=-kt+m-1
\end{equation}
where $k \geq 0, m\geq 0$ are constants. The above deceleration parameter leads to three different cases as follows:
\begin{itemize}
\item $q=-1,$ for $ k=0, m=0$
\item $q=m-1,$ for $ k=0, m>0$
\item $q=-kt+m-1,$ for $ k>0, m\geq 0$
\end{itemize}
For $q>0$ the universe exhibit decelerating expansion, constant rate of expansion for $q=0$, accelerating expansion if $-1<q<0$ (also known as power-law expansion), de Sitter expansion for $q=-1$ (also known as exponential expansion) and super exponential for $q<-1$. The first two cases i.e. for $k=0$ correspond to Berman's law of constant deceleration parameter \cite{berman83}. Therefore, only the last case for $k>0$ renders a LVDP, which is compatible with the observational data of modern cosmology.

\begin{figure}[h!]
\centering
\includegraphics[width=75mm]{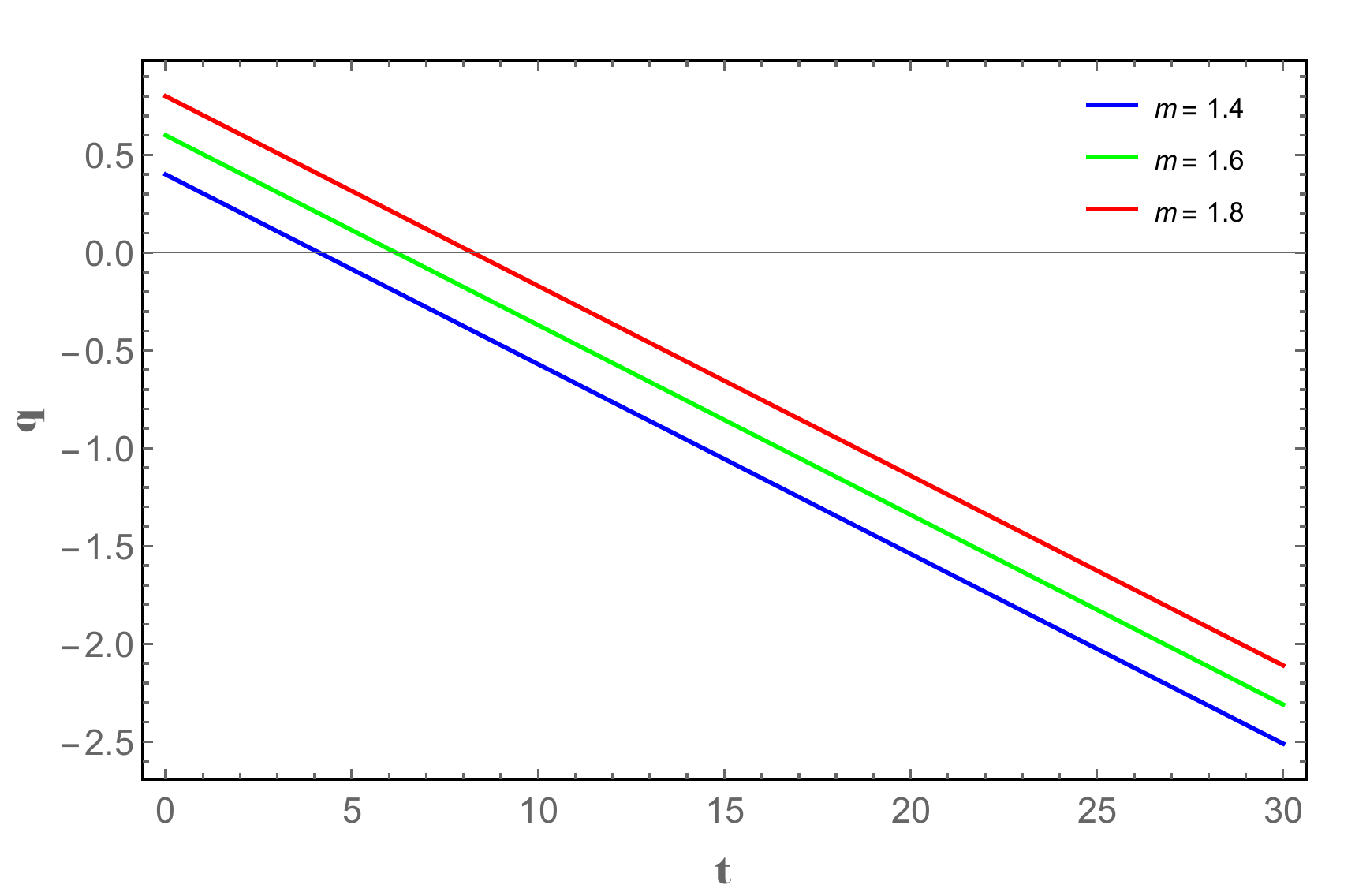}
\caption{Variation of deceleration parameter against time with $k=0.097$ and different $m$}\label{fig1}
\end{figure}

Figure 1 shows that the universe undergoes early deceleration and present acceleration. The universe begins with decelerating expansion for  $q=m-1>0$ and enters to accelerating phase at $t=\frac{m-1}{k}$. The universe enters to the acceleration phase at $t\approx4.1, 6.2, 8.2$ and present values of deceleration parameter $q=-0.938, -0.738, -0.538$ at $t=13.798$ for $m=1.4, 1.6, 1.8$ respectively. Hence, these values are consistent with respect to the observational data. Since $q=-1$ at $t=\frac{m}{k}$ indicates that the universe experiences super exponential expansion which ends with $q=-m-1$ at $t=\frac{2m}{k}$.  One can get isotroic model at $t=\frac{2m}{k}$.\\
Solving equation (31) with $k>0$ and $m\geq0$ the scale factor $a$ is obtained as
\begin{equation}
a=c \exp\bigg[ \frac{2}{\sqrt{m^2-2l k}} \text{arctanh} \bigg(\frac{kt-m}{\sqrt{m^2-2l k}}\bigg)\bigg]
\end{equation}
where $c$, $l$ are integrating constants. Assuming the integrating constant $l=0$, the scale factor and corresponding mean Hubble parameter are obtained as
\begin{equation}
a=ce^{\frac{2}{m}\text{arctanh}\left(\frac{kt}{m}-1\right)}
\end{equation}
\begin{equation}
H=-\frac{2}{kt(t-t_{BR})}
\end{equation}
where $t_{BR}=\frac{2m}{k}$.\\
The scalar expansion $\theta$ is
\begin{equation}
\theta=-\frac{6}{kt(t-t_{BR})}
\end{equation}

\begin{figure}[h!]
\centering
\includegraphics[width=75mm]{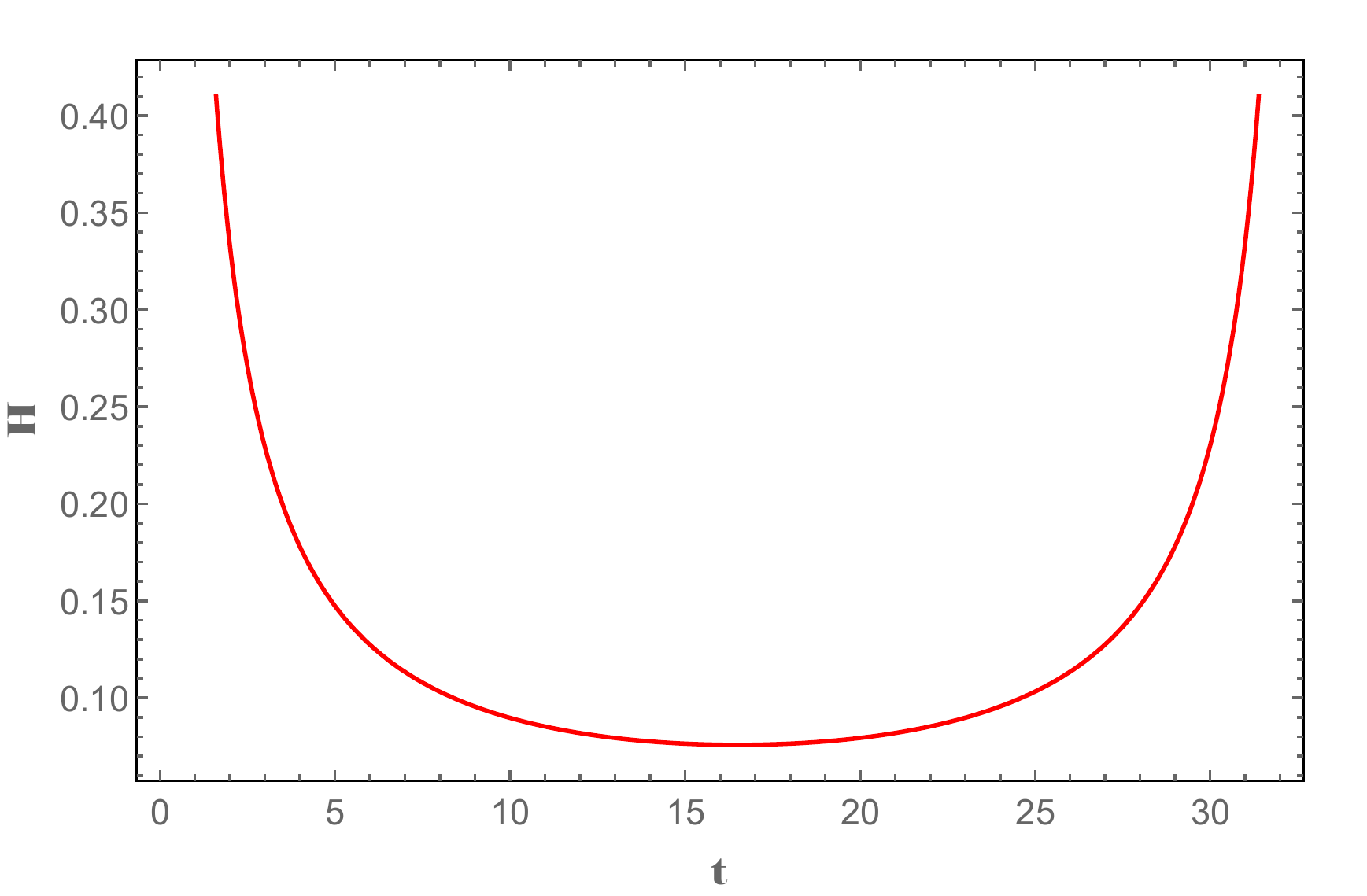}
\caption{Variation of Hubble parameter against time with $k=0.097$ and $m=1.6$}\label{fig2}
\end{figure}

The variation of Hubble parameter against time is presented in the figure 2. For the large $t$, the Hubble parameter approached towards zero i.e. $H\rightarrow 0$ when $t\rightarrow \infty$. It has singularity at $t=0 $ and $t=t_{BR}$. Hence, the
Hubble parameter and directional Hubble parameters both diverge at the beginning and at the Big Rip. The Hubble rates evolve with time in between the big bang and Big Rip i.e. the intermediate phase between initial and end of the universe. The model of the universe starts with big bang and ends with Big Rip. At transition phase the Hubble parameter becomes $H=\frac{2k}{m^2-1}$ \cite{Nojiri2010}. For an expanding universe we need the positive value of scalar expansion $\theta$. From Eq. (35) our model of the universe is expanding for $t<t_{BR}$.
\newline
The mean anisotropic parameter for the model becomes constant as given below
\begin{equation}
\mathcal{A}=\frac{1}{3}\left[\frac{6n^2-16n+4}{(n+2)^2}\right]
\end{equation}
The magnetic flux for our model becomes
\begin{equation}
h^2=\frac{3(n-1)(-kt+m-3)}{2(4\pi+\mu)(n+2)}\left(\frac{-2}{kt(t-t_R)}\right)^2
\end{equation}

\begin{figure}[h!]
\centering
\includegraphics[width=75mm]{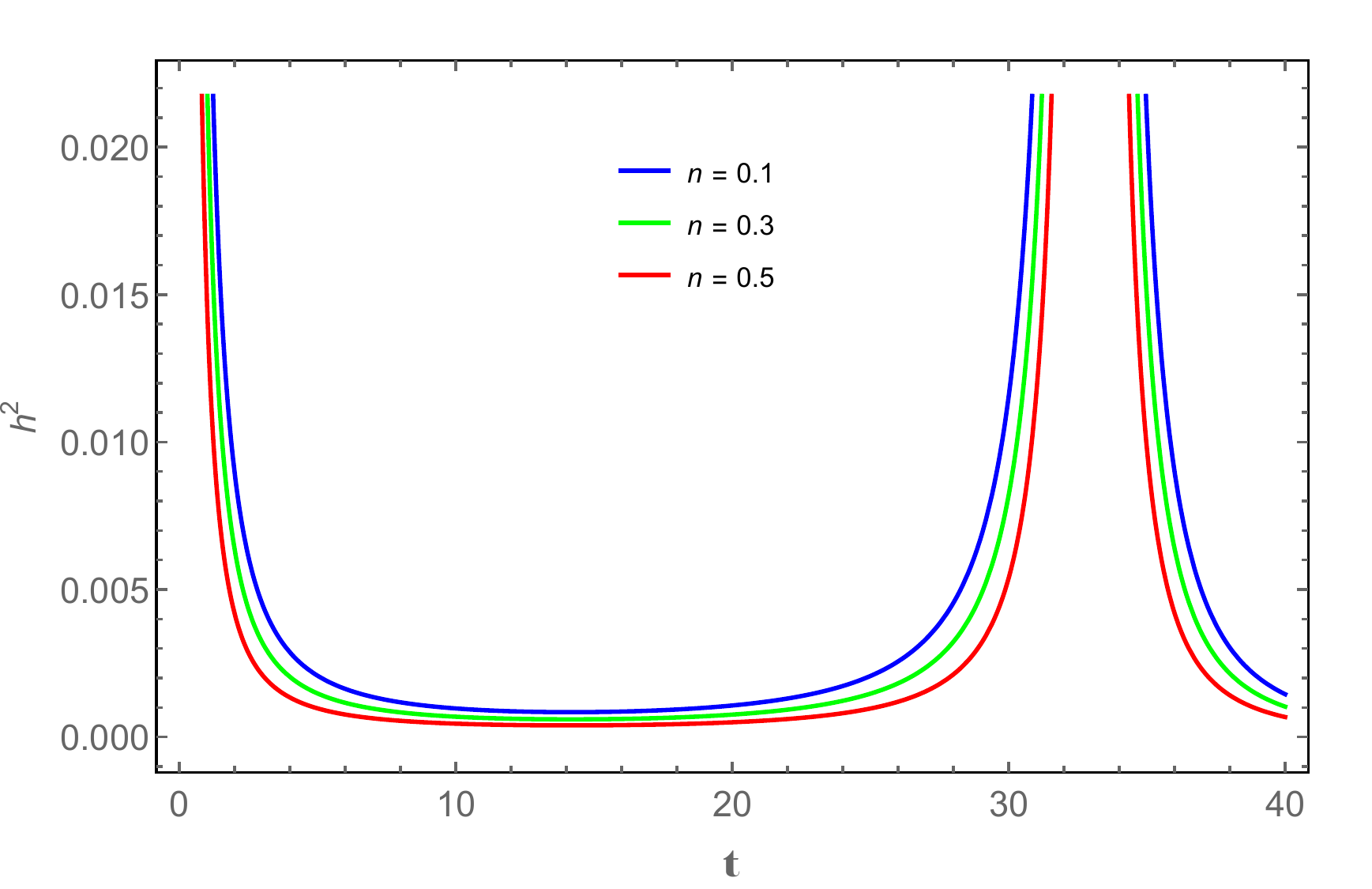}
\caption{Variation of magnetic flux $h^2$ against time with $k=0.097$, $m=1.6$, $\mu=0.1$ and different $n$}\label{fig3}
\end{figure}

Figure 3 represent the variation of magnetic flux $h^2$ against time. Here the positivity of $h^2$ demands that $n\in (0,1)$ and for $n>1$, $h^2<0$. Thus we neglect the case and variation of $h^2$ is presented for $n\in (0,1)$ with $k=0.097$, $m=1.6$ and $\mu=0.1$. Here $h^2\rightarrow 0$ when $t\rightarrow \infty$. The parameter magnetic flux has also the same singularity as that of Hubble parameter.\newline
Using the above values we obtain the energy density $\rho$ and pressure $p$ for our model as
\begin{equation}
\rho=\frac{-3}{4(4\pi+\mu)}\biggl[\frac{9(n-1)}{(n+2)^2}+\frac{3[3-3n+(-kt+m)n]}{(n+2)}\biggr]H^2+B_c
\end{equation}
\begin{equation}
p=\frac{-1}{4(4\pi+\mu)}\biggl[\frac{9(n-1)}{(n+2)^2}+\frac{3[3-3n+(-kt+m)n]}{(n+2)}\biggr]H^2-B_c
\end{equation}

\begin{figure}[h!]
\minipage{0.32\textwidth}
  \includegraphics[width=45mm]{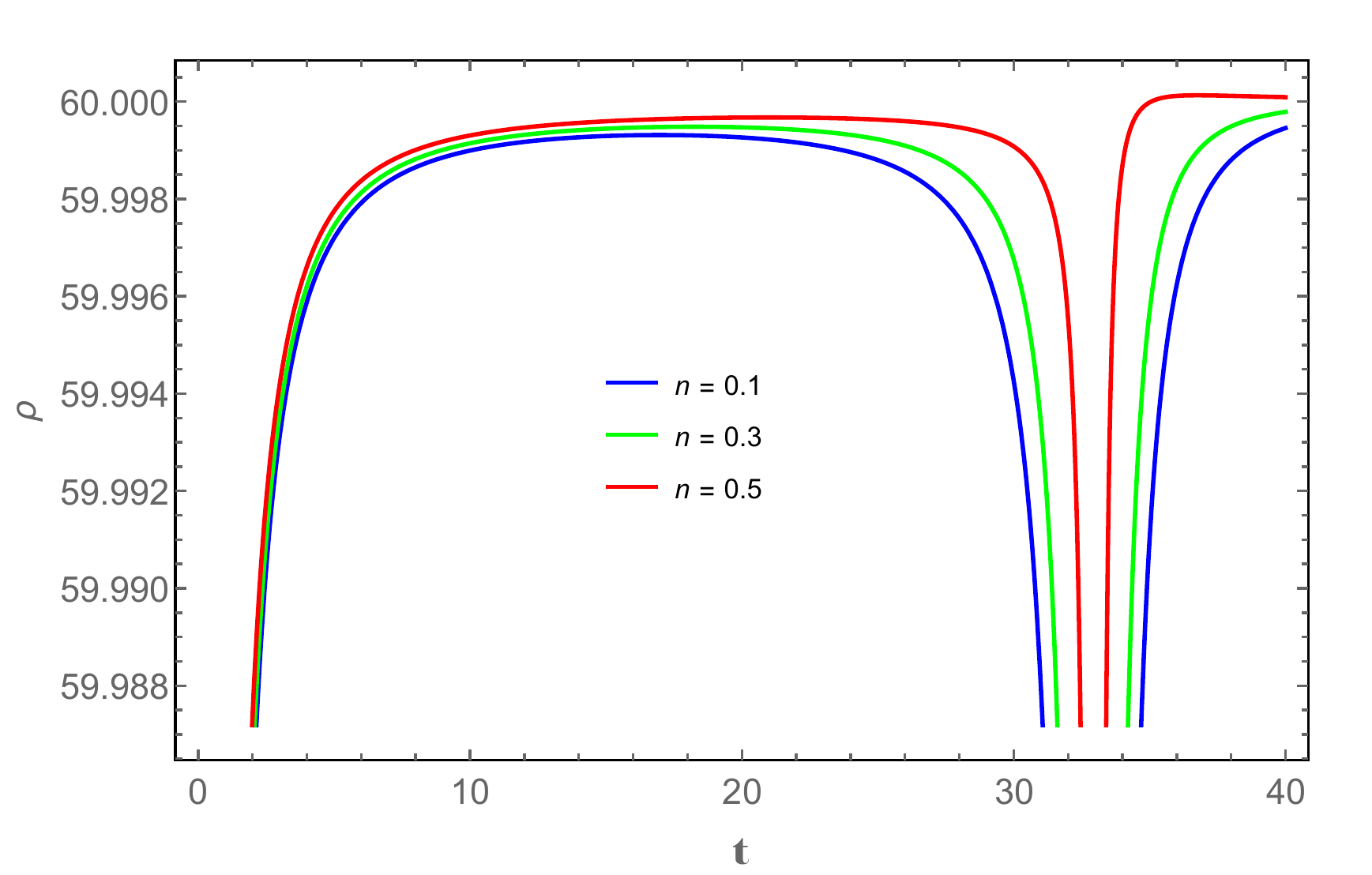}
\endminipage\hfill
\minipage{0.32\textwidth}
  \includegraphics[width=45mm]{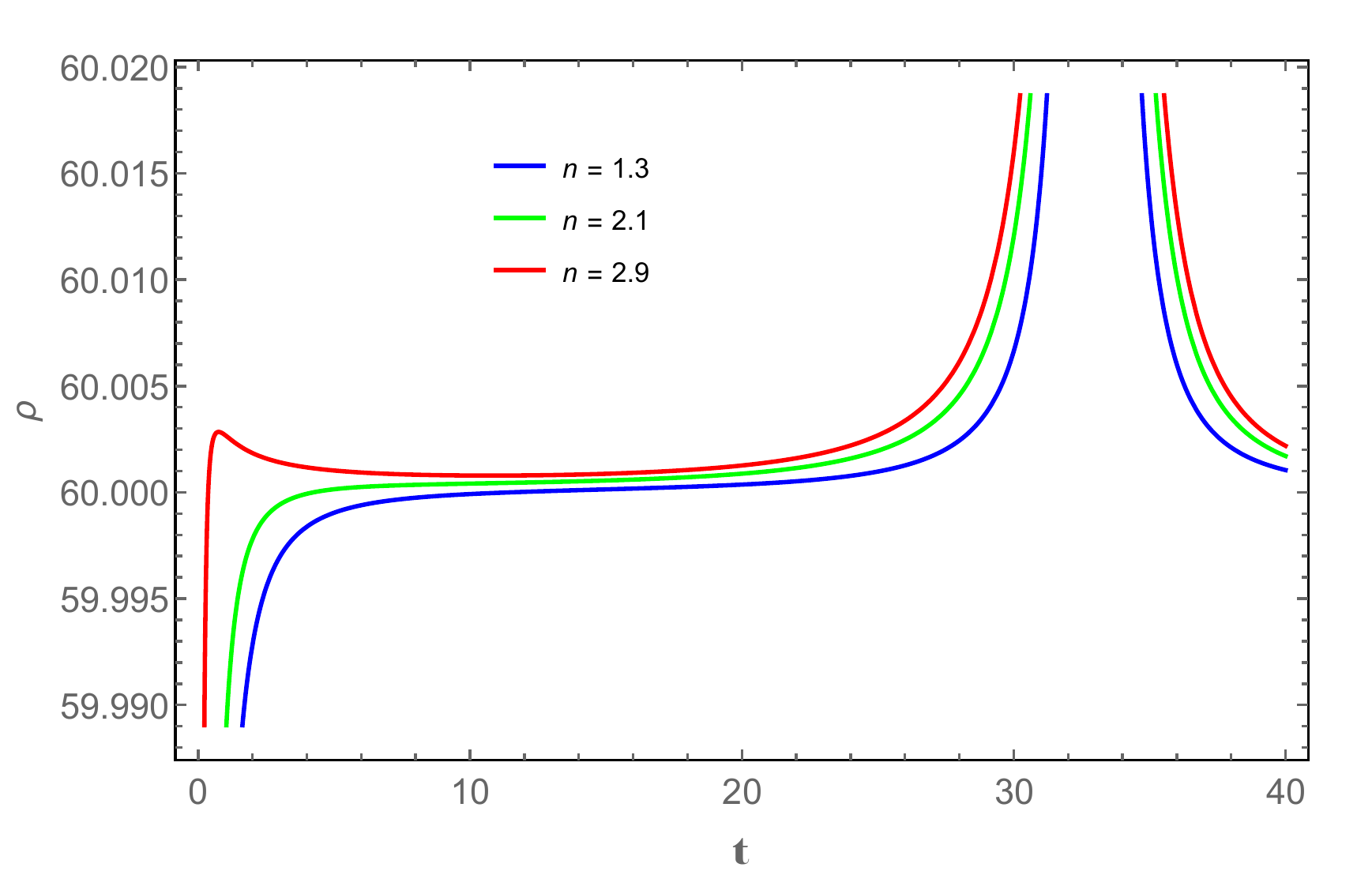}
\endminipage\hfill
\minipage{0.32\textwidth}%
  \includegraphics[width=45mm]{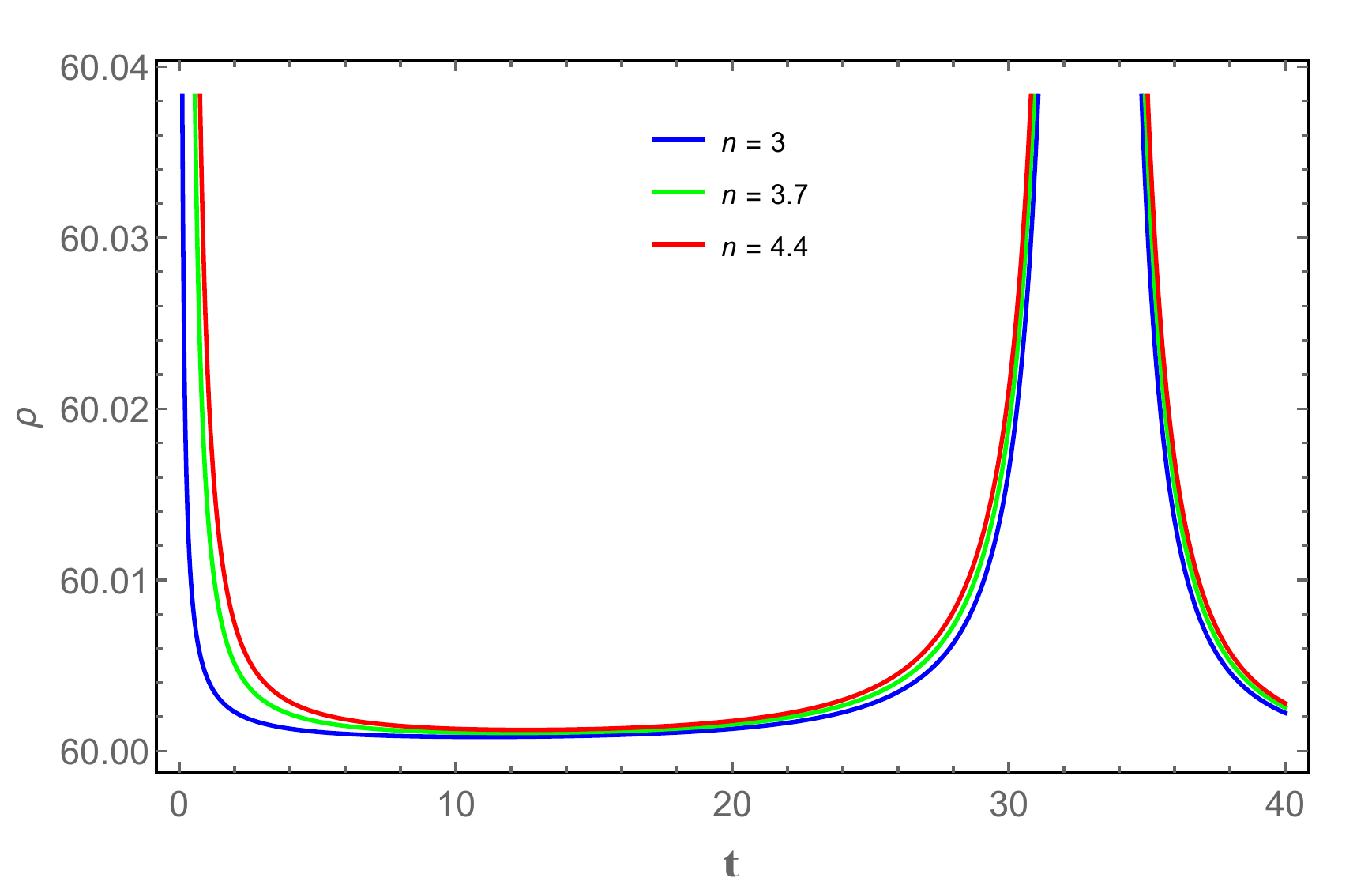}
\endminipage
\caption{Variation of energy density against time with $k=0.097$, $m=1.6$, $\mu=0.1$, $B_c=60$ and different $n$ i.e. $n\in (0,0.5]$, $n\in [0.6,3)$ and $n\in [3,\infty)$}\label{fig4}
\end{figure}

The profile of energy density against time is presented in the figure 4. From equation (38), one can observe that, $\rho\rightarrow B_c$ when $t\rightarrow \infty$. The energy density approaches to $B_c$ in different ways for different interval of $n$, which is presented in the figure 4. As time increases the energy density of the fluid diverges very fast leading to Big Rip singularity at finite time $t_R=\frac{2m}{k}$.

\begin{figure}[ht!]
\minipage{0.32\textwidth}
  \includegraphics[width=45mm]{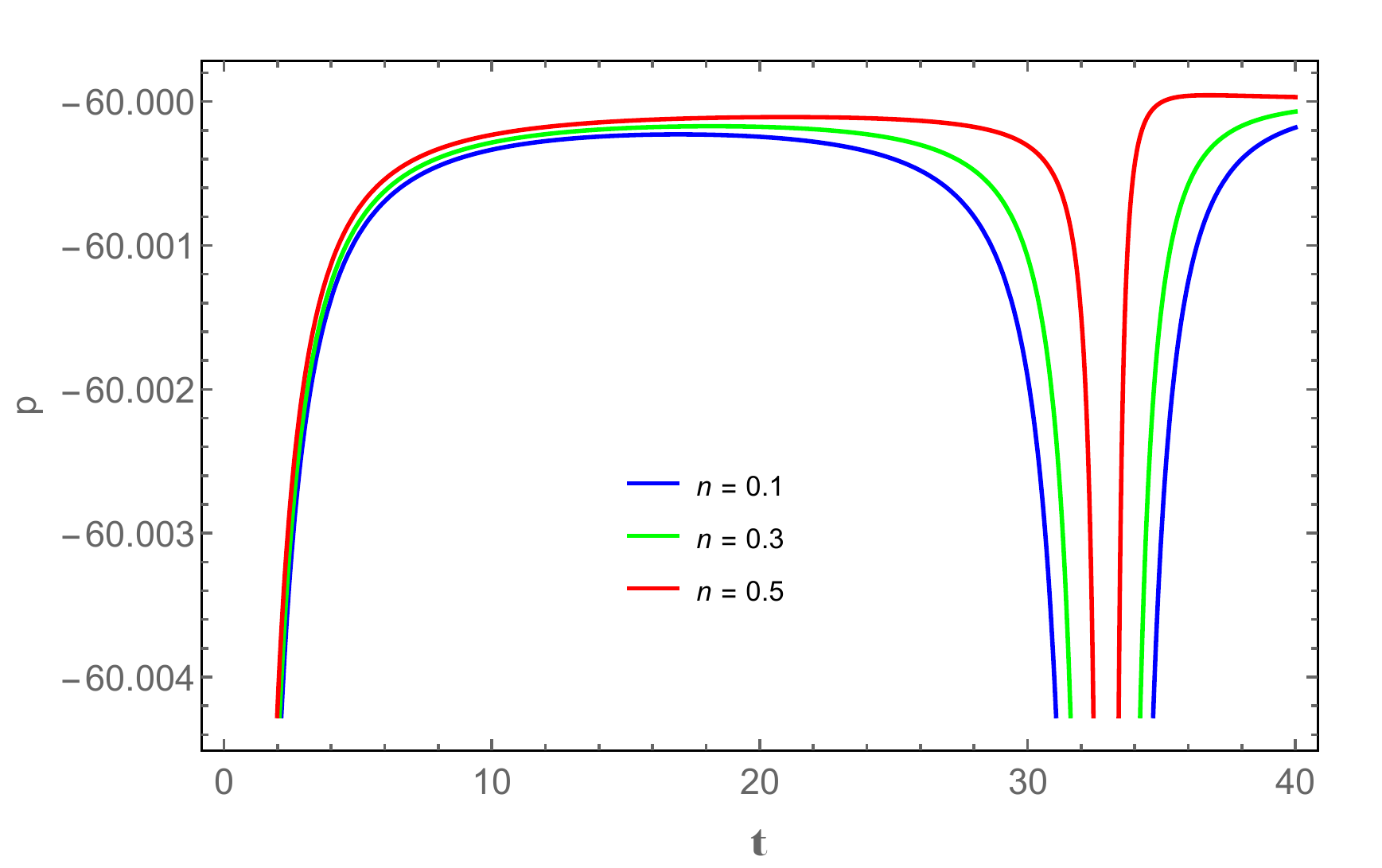}
\endminipage\hfill
\minipage{0.32\textwidth}
  \includegraphics[width=45mm]{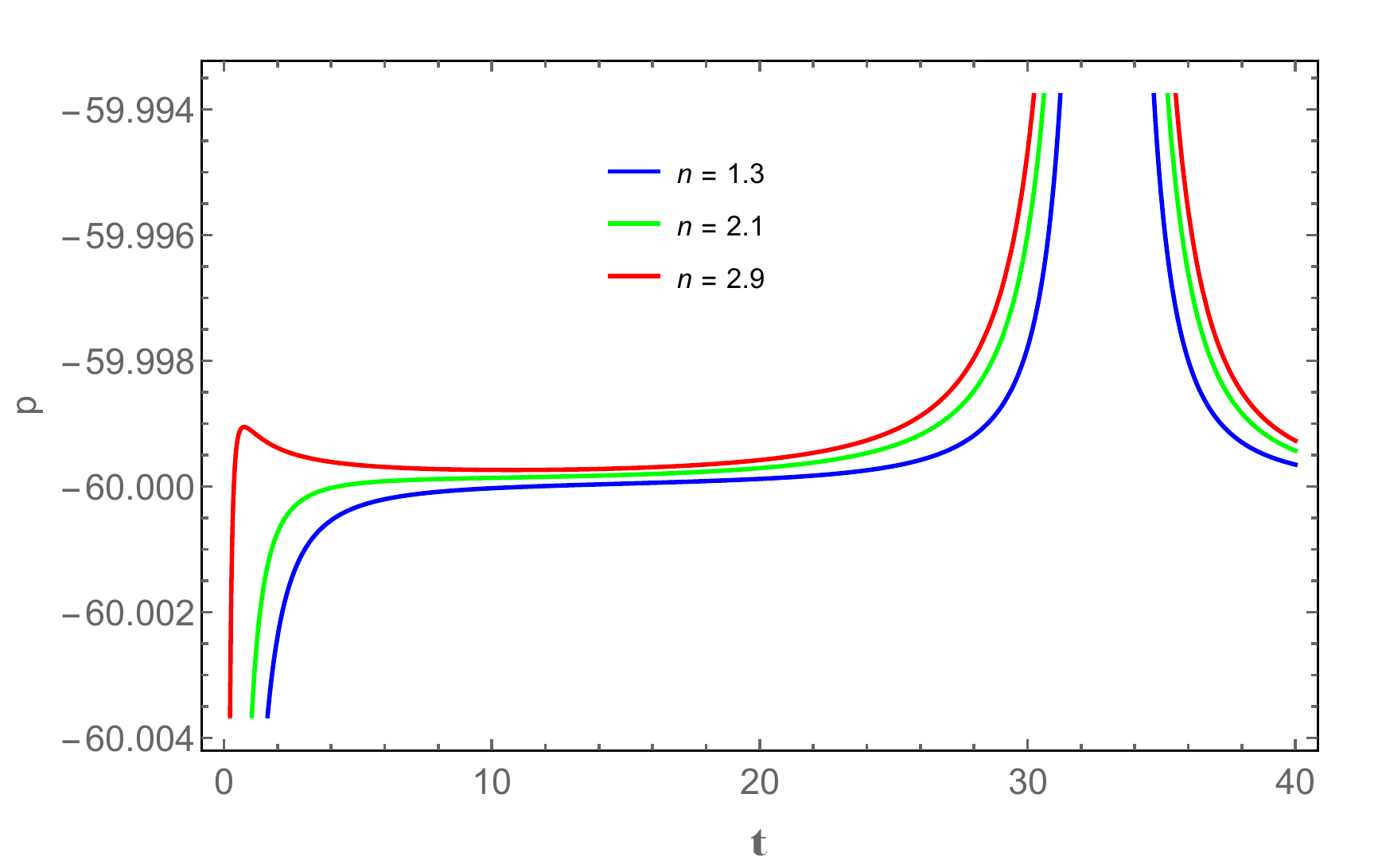}
\endminipage\hfill
\minipage{0.32\textwidth}%
  \includegraphics[width=45mm]{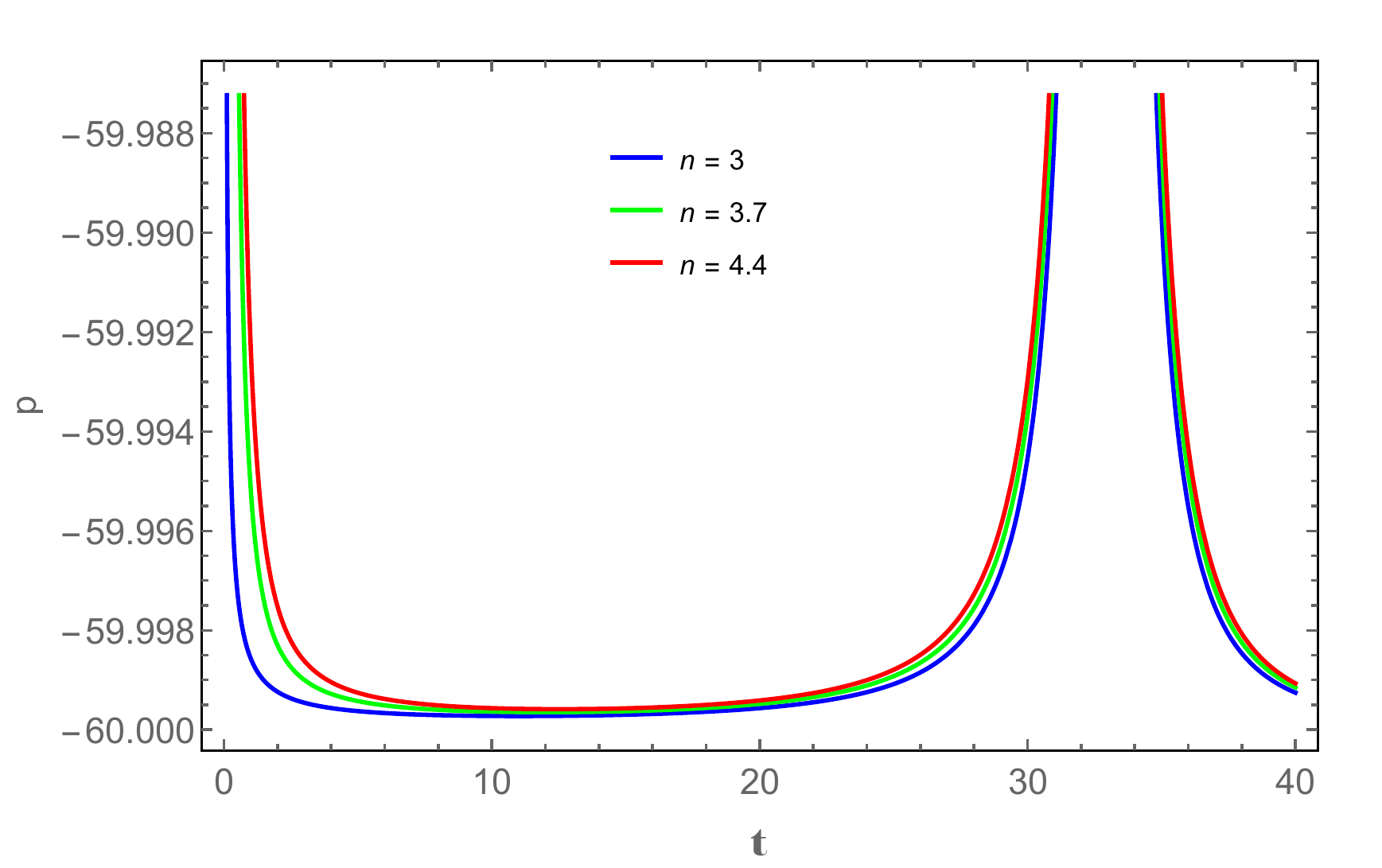}
\endminipage
\caption{Variation of pressure against time with $k=0.097$, $m=1.6$, $\mu=0.1$, $B_c=60$ and different $n$ i.e. $n\in (0,0.5]$, $n\in [0.6,3)$ and $n\in [3,\infty)$}\label{fig5}
\end{figure}

Consequently, our EoS $\omega$ is as follows
\begin{equation}
\omega=\frac{-{B_c}-\frac{\frac{3 (n (m-k t)-3 n+3)}{n+2}+\frac{9 (n-1)}{(n+2)^2}}{k^2 (\mu +4 \pi ) t^2 \left(t-t_R \right)^2}}{{B_c}-\frac{3 \left(\frac{3 (n (m-k t)-3 n+3)}{n+2}+\frac{9 (n-1)}{(n+2)^2}\right)}{k^2 (\mu +4 \pi ) t^2 \left(t-t_R \right)^2}}
\end{equation}

\begin{figure}[ht!]
\minipage{0.32\textwidth}
  \includegraphics[width=45mm]{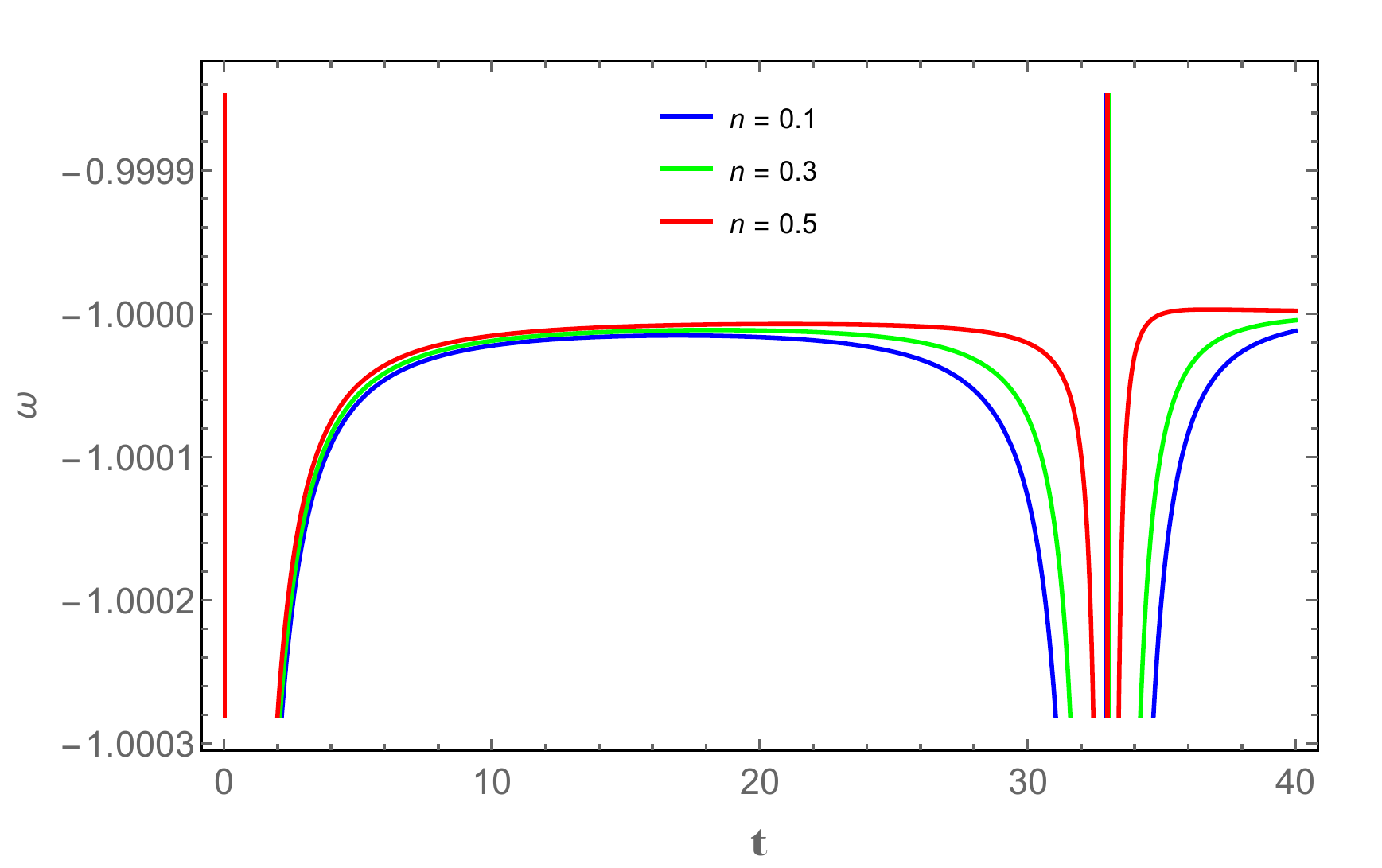}
\endminipage\hfill
\minipage{0.32\textwidth}
  \includegraphics[width=45mm]{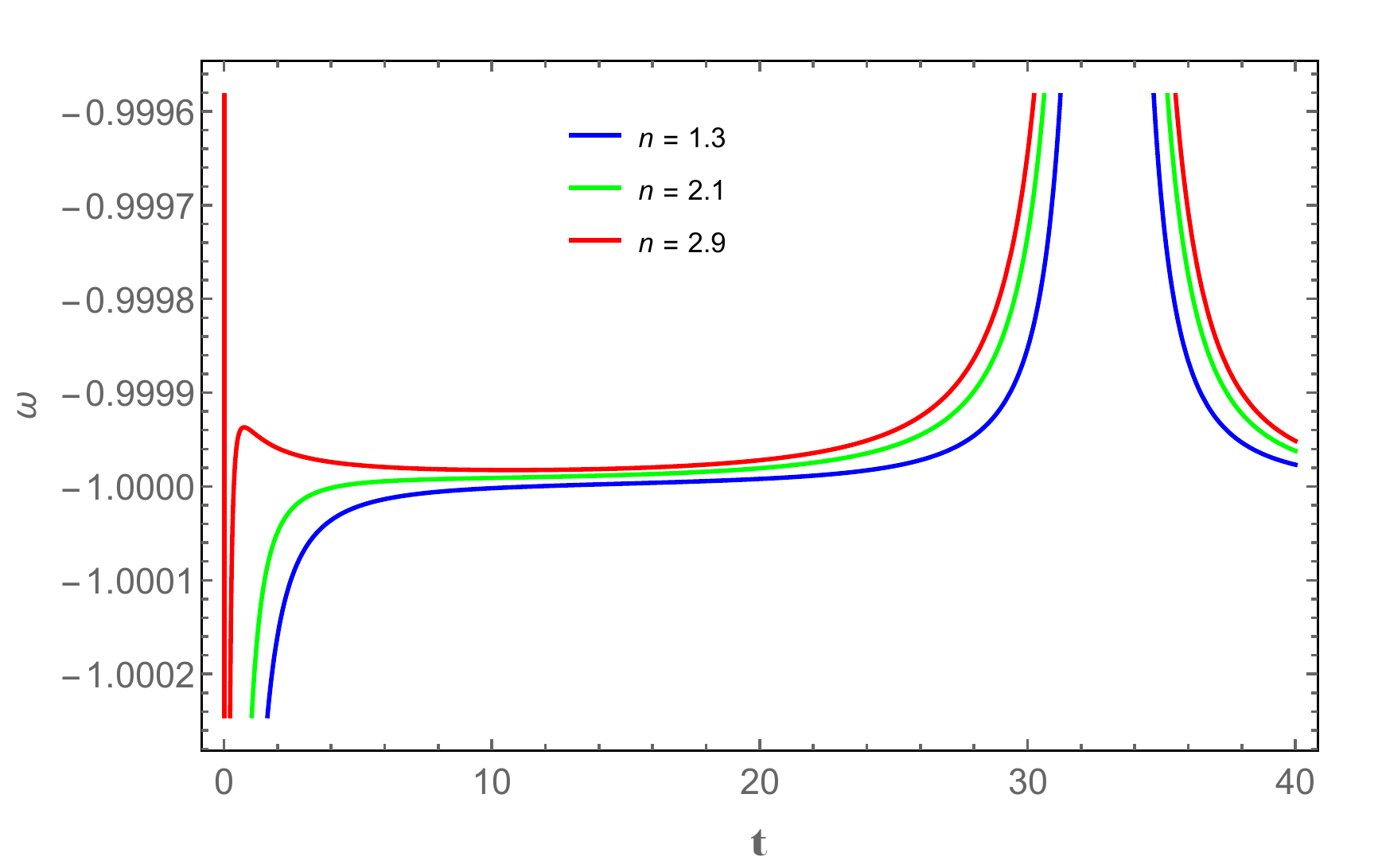}
\endminipage\hfill
\minipage{0.32\textwidth}%
  \includegraphics[width=45mm]{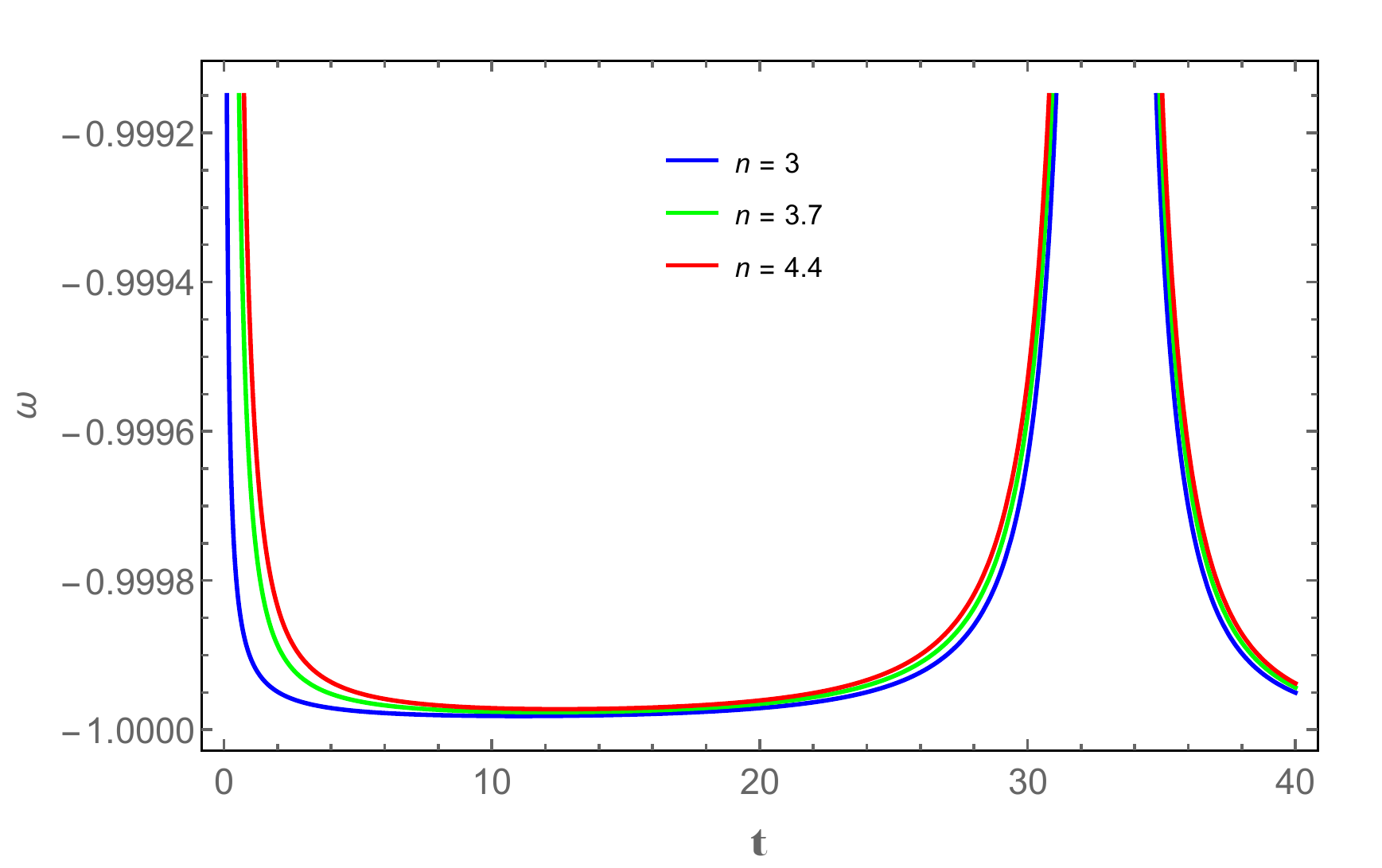}
\endminipage
\caption{Variation of EoS $\omega$ against time with $k=0.097$, $m=1.6$, $\mu=0.1$, $B_c=60$ and different $n$ i.e. $n\in (0,0.5]$, $n\in [0.6,3)$ and $n\in [3,\infty)$}\label{fig6}
\end{figure}

The pressure profile has also the same singularity as that of Hubble parameter, which is noticed from the figure 5. From equation (39) one can observe that, $p\rightarrow -B_c$ when $t\rightarrow \infty$. Pressure is negative here and it approaches to $-B_c$ in different way for different interval of $n$ (See figure 5). The cosmological constant is negative, which follow the observational data. Figure 6 represents the variation of EoS parameter against time. From equation (40) one can observe that, $\omega\rightarrow -1$ when $t\rightarrow \infty$. The EoS parameter approaches to $-1$ in different way for different interval of $n$ (See figure 6). It follows the recent observational data. The parameter EoS has also the same singularity as that of Hubble parameter i.e. at the initial phase and at the Big Rip $t_{BR}$.\newline
The cosmological constant is
\begin{multline}
\Lambda=\biggl[\frac{3[(12n\pi+3n\mu-n^2 \mu+24\pi+10\mu)(-kt+m-1)]}{2(4\pi+\mu)(n+2)^2}\\+\frac{(-26\mu+18n\mu+6n^2 \mu-76\pi)}{2(4\pi+\mu)(n+2)^2}\biggr]H^2-(8\pi+4\mu)B_c
\end{multline}

\begin{figure}[ht!]
\minipage{0.32\textwidth}
  \includegraphics[width=45mm]{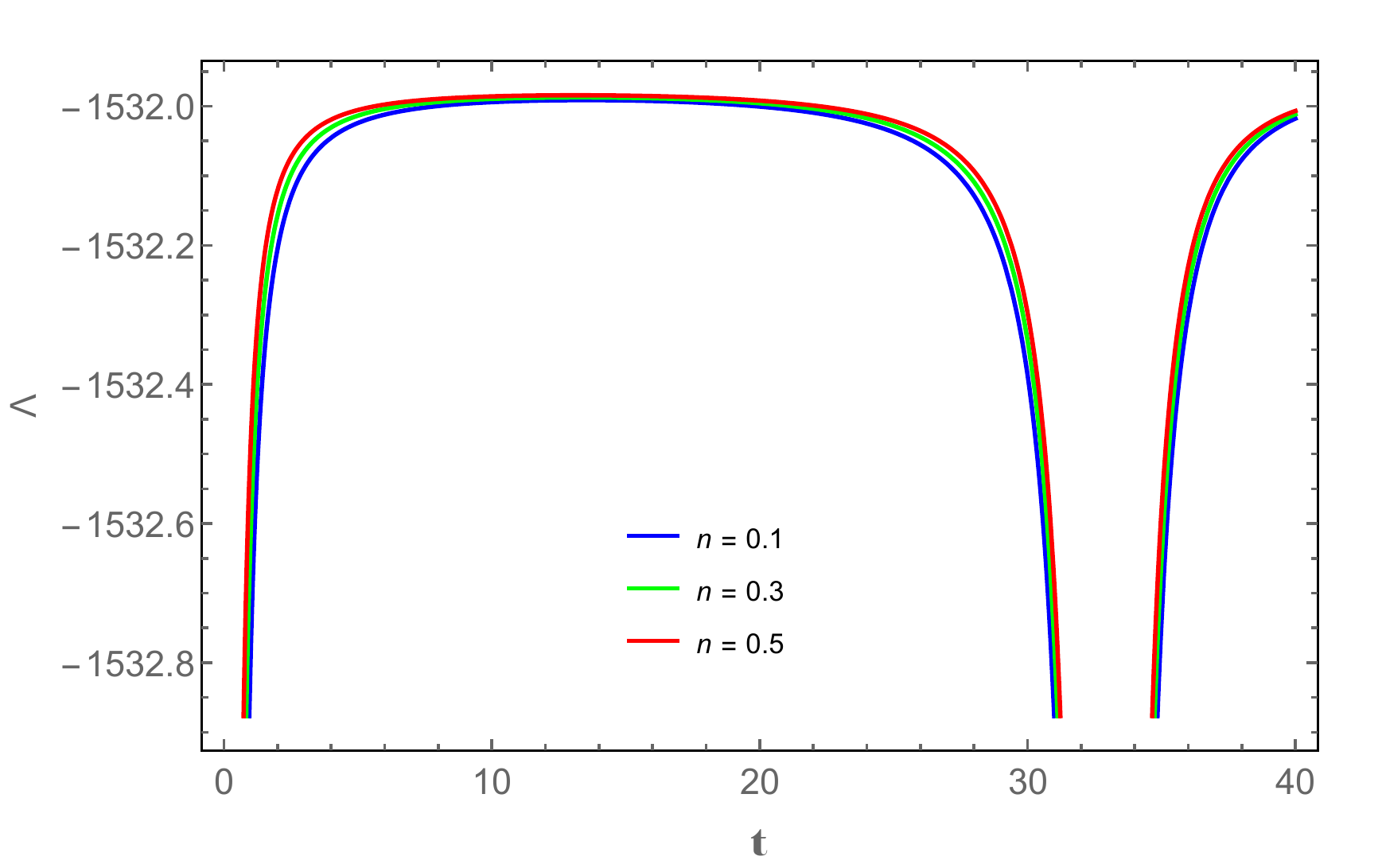}
\endminipage\hfill
\minipage{0.32\textwidth}
  \includegraphics[width=45mm]{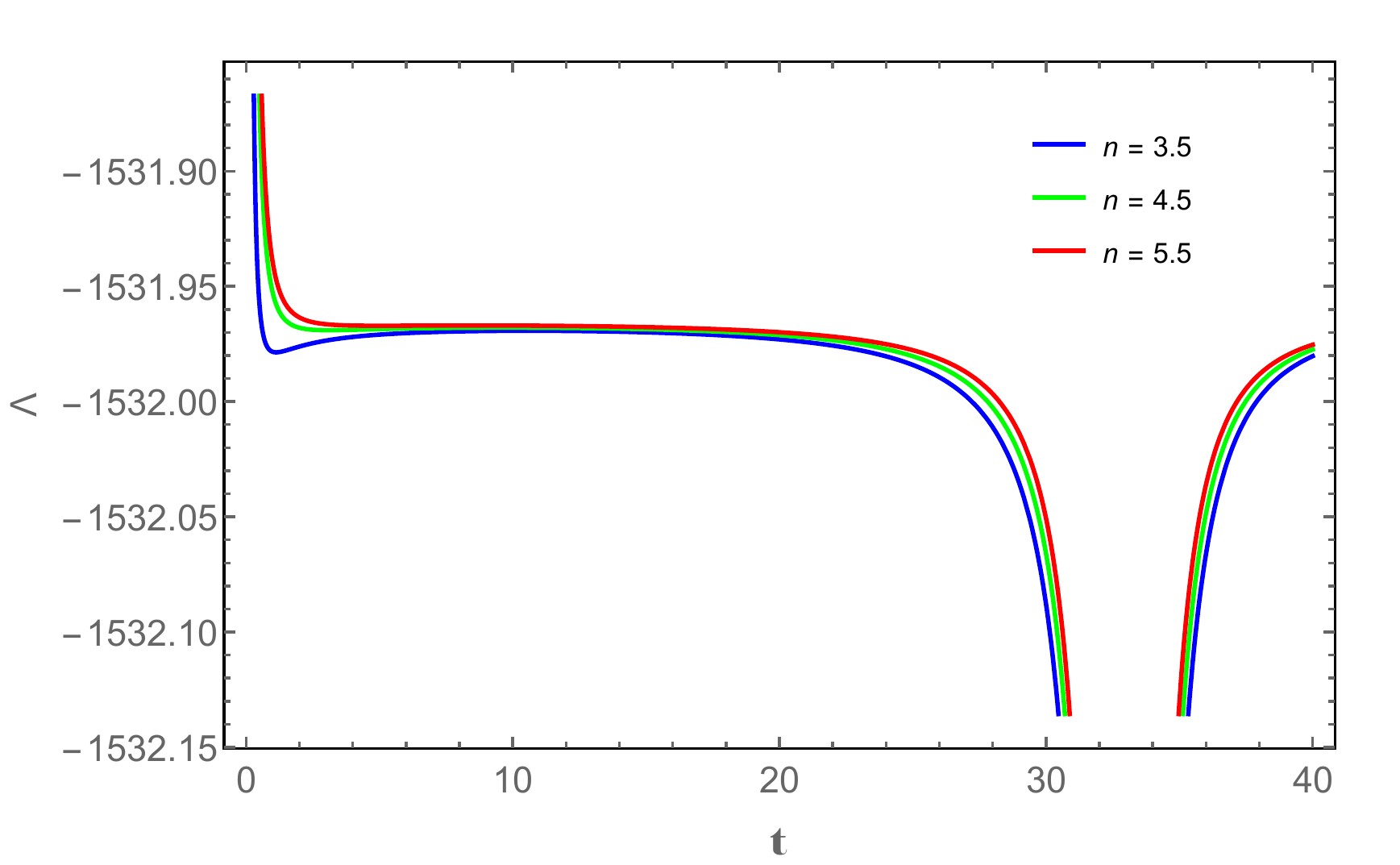}
\endminipage\hfill
\minipage{0.32\textwidth}%
  \includegraphics[width=45mm]{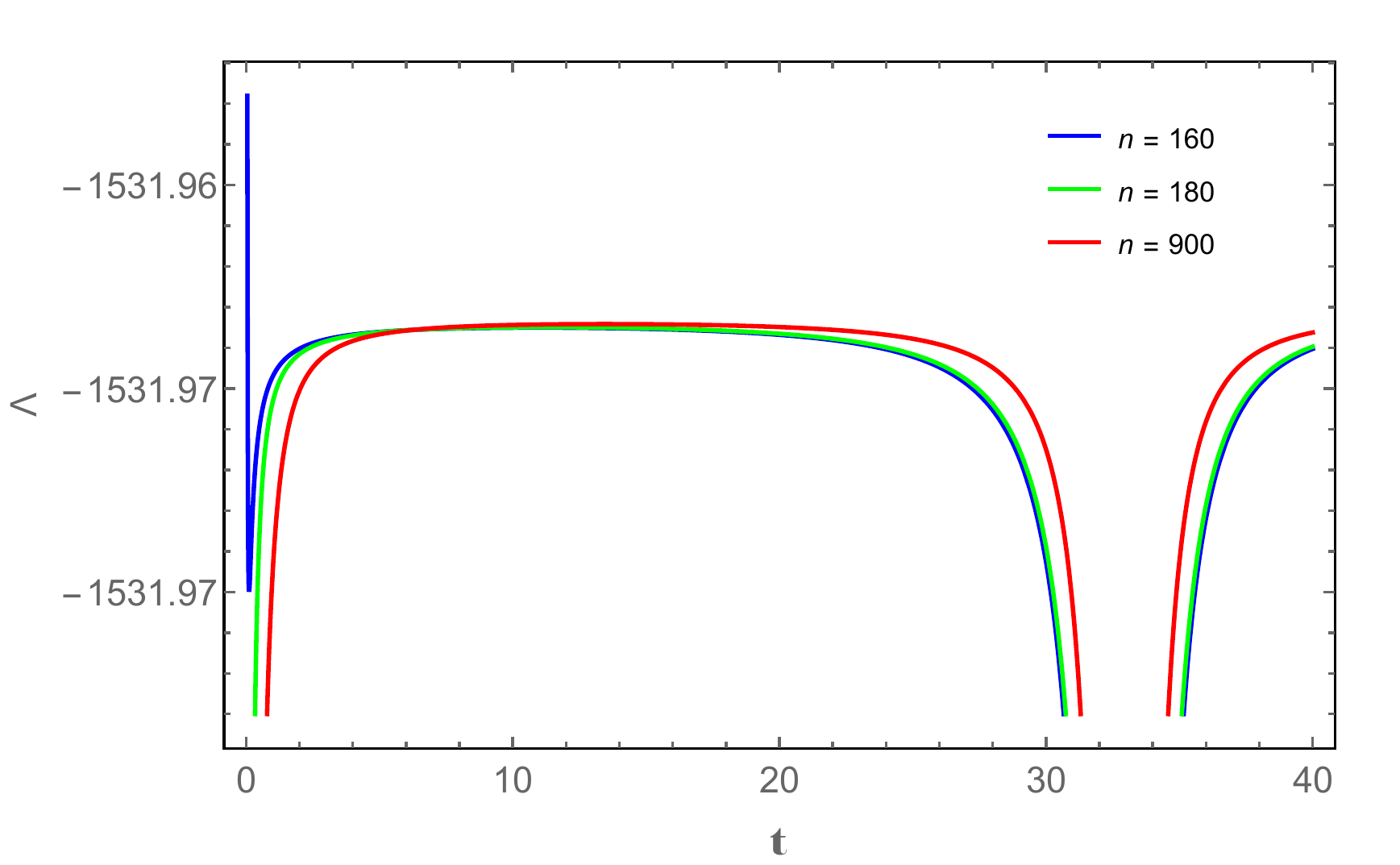}
\endminipage
\caption{Variation of cosmological constant against time with $k=0.097$, $m=1.6$, $\mu=0.1$, $B_c=60$  and different $n$ i.e. $n\in (0,3]$, $n\in (3,160]$ and $n\in [160,\infty)$}\label{fig7}
\end{figure}

The profile of cosmological constant against time is presented in the figure 7. From equation (41), one can observe that, $\Lambda\rightarrow -(8\pi+4\mu)B_c$, when $t\rightarrow \infty$. For different interval of $n$, $\Lambda$ approaches to $-(8\pi+4\mu)B_c$ in different ways, which can be noticed from figure 7. The parameter $\Lambda$ has also the same singularity as that of Hubble parameter.

The Ricci scalar $R$ and the trace of energy momentum tensor $T$ are obtained as
\begin{equation}
R = -\bigg[ 2\frac{\ddot{A}}{A}+4\frac{\ddot{B}}{B}+4\frac{\dot{A}\dot{B}}{AB}+2\frac{\dot{B}^2}{B^2}\bigg]
=-\bigg[\frac{3n(2n+4)}{n+2}(-1-q)+\frac{9(2n^2+4n+6)}{(n+2)^2}\bigg]H^2
\end{equation}
and
\begin{equation}
T=\rho-3p+2h^2=4B_c +\frac{6(n-1)(q-2)}{2(4\pi+\mu)(n+2)}H^2
\end{equation}
Using equations (42) and (43), the function $f(R,T)$ can be obtained as
\begin{multline}
f(R,T)=8\mu B_c+ \frac{[(4\pi+\mu)(6n^2+12n)+6\mu(n-1)](-kt+m-1)}{(n+2)(4\pi+\mu)}\\+
\frac{6n^3+6n^2-12n-54}{(n+2)^2}-\frac{12\mu(n-1)}{(n+2)(4\pi+\mu)}
\end{multline}

\begin{figure}[h!]
\centering
\includegraphics[width=75mm]{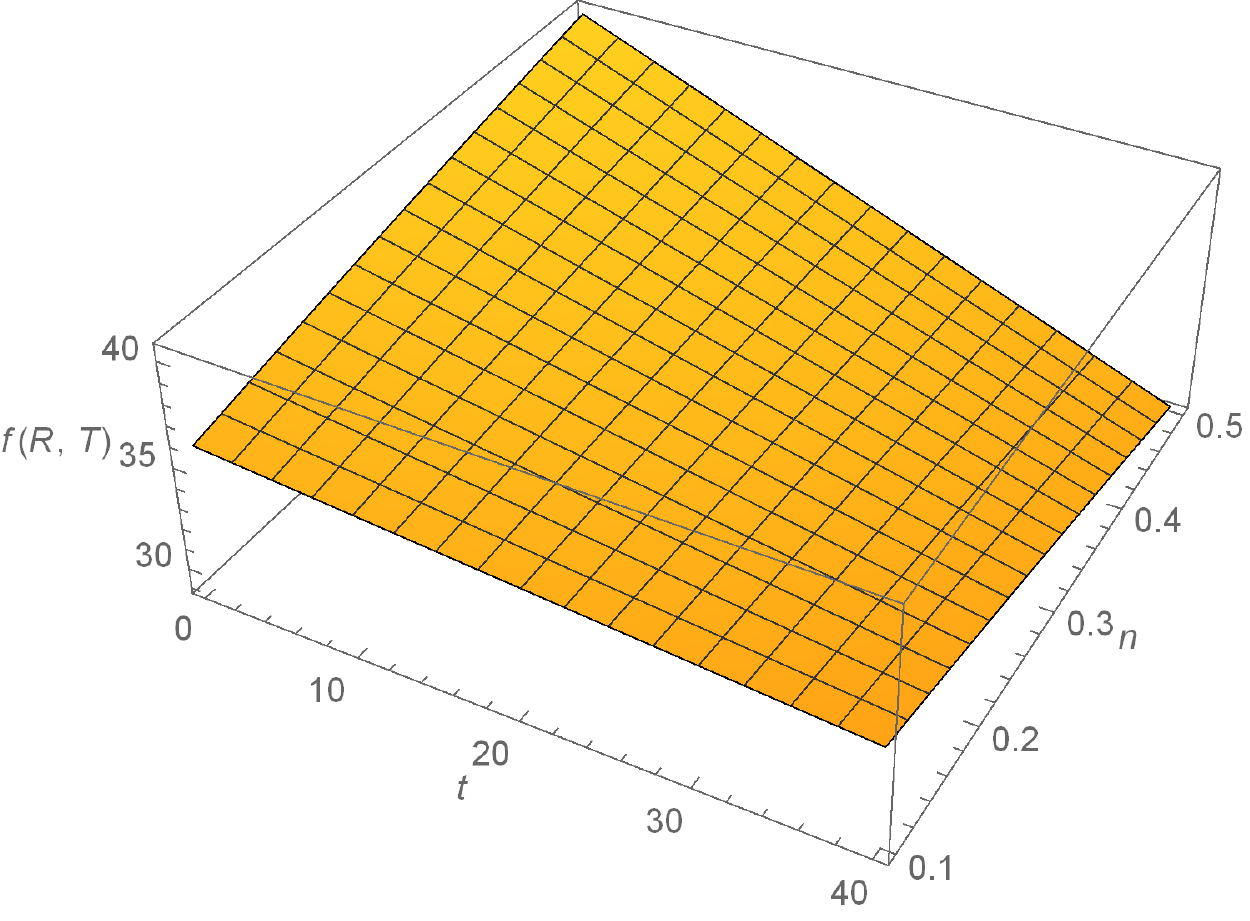}
\caption{Behaviour of $f(R,T)$ versus $t$ and $n$ with $\mu=0.1$, $B_C=60$, $m=1.6$ and $k=0.097$ respectively.}\label{fig8}
\end{figure}
Figure 8 represents the behaviour of $f(R,T)$ for this model.
\section{The dynamics of the model}
The scale factor $a$ in terms of redshift parameter $z$ is written as
\begin{equation}
a=\frac{a_0}{1+z}
\end{equation}
where, $a_0$ is the present scale factor.\\
From Eqs. (33) and (45) we get
\begin{equation}
z=\left(\frac{e^{\frac{2}{m}\text{arctanh}\left(\frac{kt_0}{m}-1\right)}}{e^{\frac{2}{m}\text{arctanh}\left(\frac{kt}{m}-1\right)}}\right)^\frac{1}{m}
\end{equation}
Using this Eq. (47), the Hubble parameter in terms of redshift is
\begin{equation}
H=H_0(1+z)^m\left(\frac{t_0}{t}\right)^2
\end{equation}
where, $h_0$ is the value of present Hubble's parameter.
The distance modulus $\mu(z)$ is defines as
\begin{equation}
\mu(z)=5 \log d_L+25
\end{equation}
where, $d_L$ is the luminosity distance and defined as
\begin{equation}
d_L=r_1(1+z)a_0
\end{equation}
where
\begin{multline}
r_1=\int_t^{t_0}\frac{dt}{a}= \int_t^{t_0}\frac{dt}{e^{\frac{2}{m}\text{arctanh}\left(\frac{kt}{m}-1\right)}}\\=\frac{1}{c_1(9m-1)}\left\{mt\left(\frac{2m}{kt}\right)^\frac{1}{m}\times_2F_1\left[1-\frac{1}{m},-\frac{1}{m},2-\frac{1}{m},\frac{kt}{2m}\right]\right\}_t^{t_0}
\end{multline}
here, $r_1$ is a function of time $t$ at which the light we see at present time $t_0$ was emitted by the object.
The deceleration parameter $q$ in terms of $z$ is
\begin{equation}
q=2m-1-m\ \text{tanh}\left[\frac{m}{2}\text{ln}(z+1)-\text{arctanh}\left(\frac{1+q_0}{m}-2\right)\right].
\end{equation}
where, $q_0=q_{z=0}$ is the present deceleration parameter.

\begin{figure*}[h!]
\centering
\begin{minipage}[b]{.4\textwidth}
\includegraphics[width=60mm]{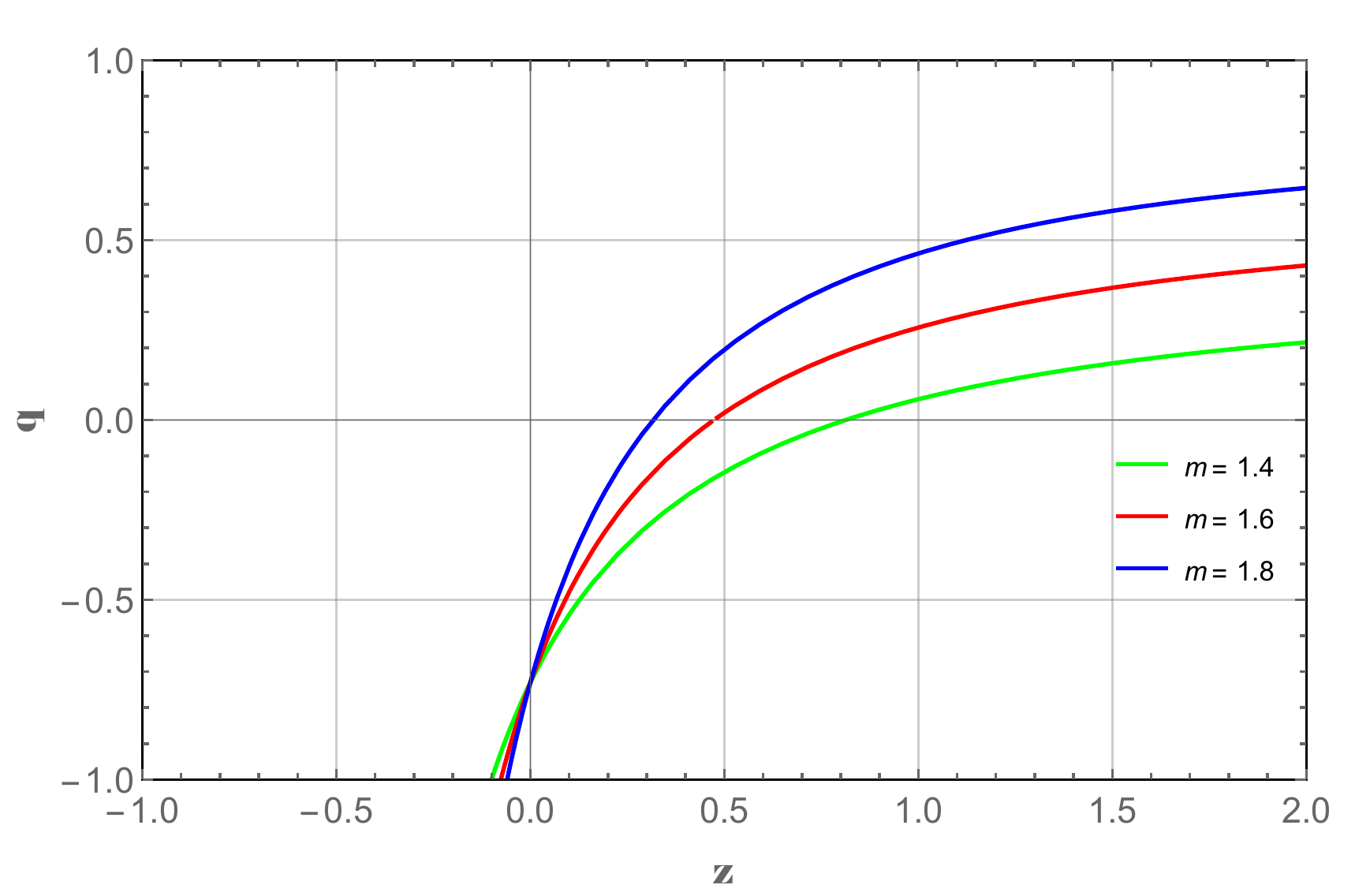}
\caption{Variation of $q$ versus $z$ with $q_0=-0.73$ and different $m$.}\label{label-9}
\end{minipage}\qquad
\begin{minipage}[b]{.4\textwidth}
\includegraphics[width=60mm]{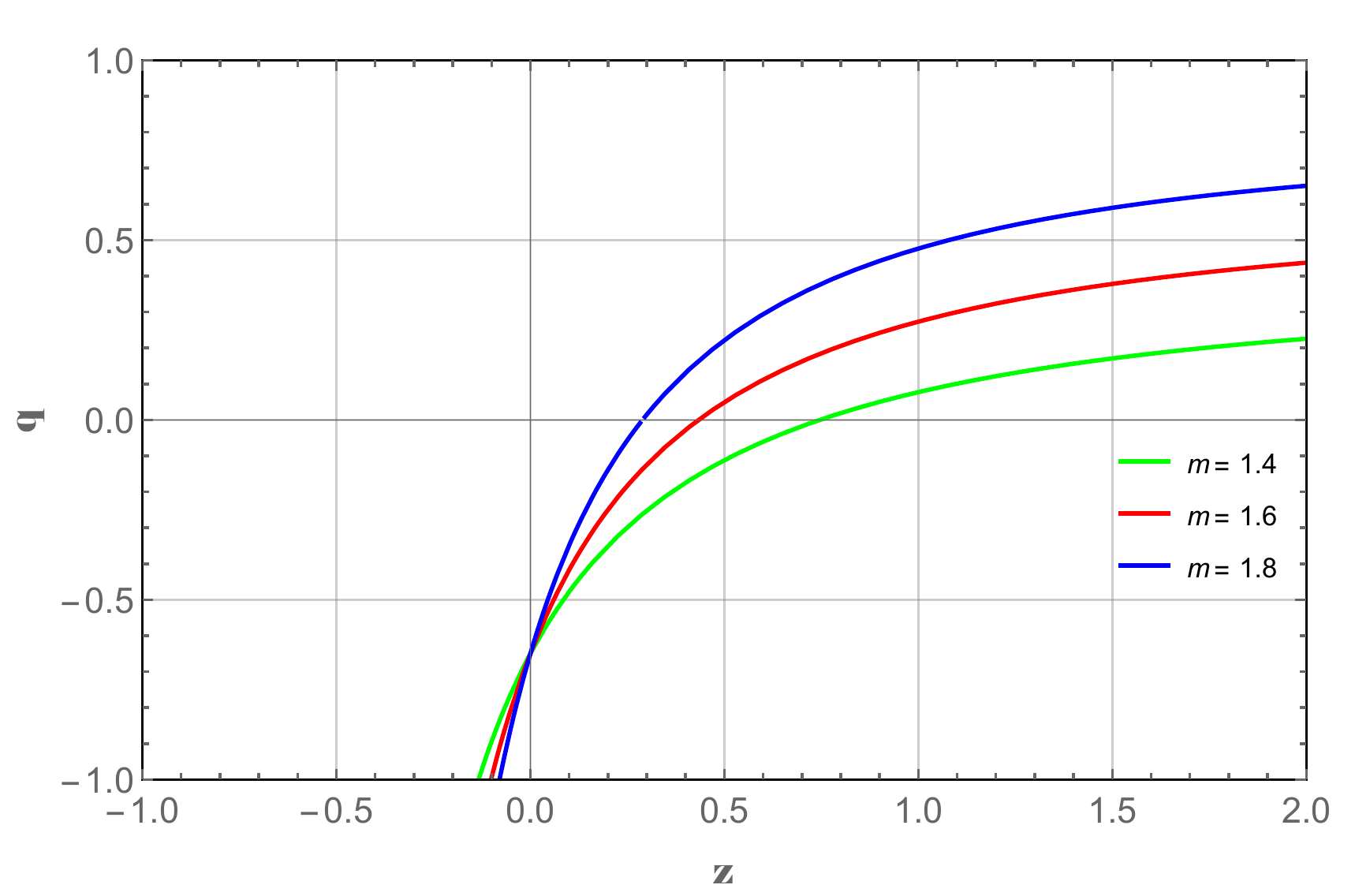}
\caption{Variation of $q$ versus $z$ with $q_0=-0.65$ and different $m$.}\label{label-10}
\end{minipage}
\end{figure*}

Here, in the figures 9 and 10 the values of $q_{z=0}=-0.73$ and $q_{z=0}=-0.65$ are considered as per the kinematic data analysis of Cuhna \cite{Cunha09} and Li et al. \cite{Li11} respectively. Again, for $q_{z=0}=-0.73$ the transition redshift $z_{tr}$ from deceleration to acceleration is taking place at $z_{tr}=0.82,\ \ 0.48,\ \ 0.327$ corresponding to $m=1.4,\ \ m=1.6,\ \ m=1.8$ respectively. Similarly, in the right figure for $q_{z=0}=-0.65$ the transition redshift values are $z_{tr}=0.75,\ \ 0.44,\ \ 0.29$ corresponding to $m=1.4,\ \ m=1.6,\ \ m=1.8$ respectively. Our $z_{tr}$ values of transition redshift fit with the observational data \cite{Capozziello14,Capozziello15,Farooq17}.\\
Here, we intends to compare our model with $\Lambda$CDM model by plotting the evolution trajectories of the $\{q,j\}$ and $\{j,s\}$. The jerk parameter $j$ has the value
\begin{equation}
j=\frac{a^2\dddot{a}}{\dot{a}^3}=\frac{3 k^2 t^2}{2}-3 m (k t+1)+3 k t+2 m^2+1
\end{equation}
The $s$ parameter is defined as \cite{Akarsu14}
\begin{equation}
s=\frac{j-1}{3(q-1)}=\frac{-6 m (k t+1)+3 k t (k t+2)+4 m^2}{6 (-k t+m-2)}
\end{equation}

\begin{figure*}[h!]
\centering
\begin{minipage}[b]{.4\textwidth}
\includegraphics[width=60mm]{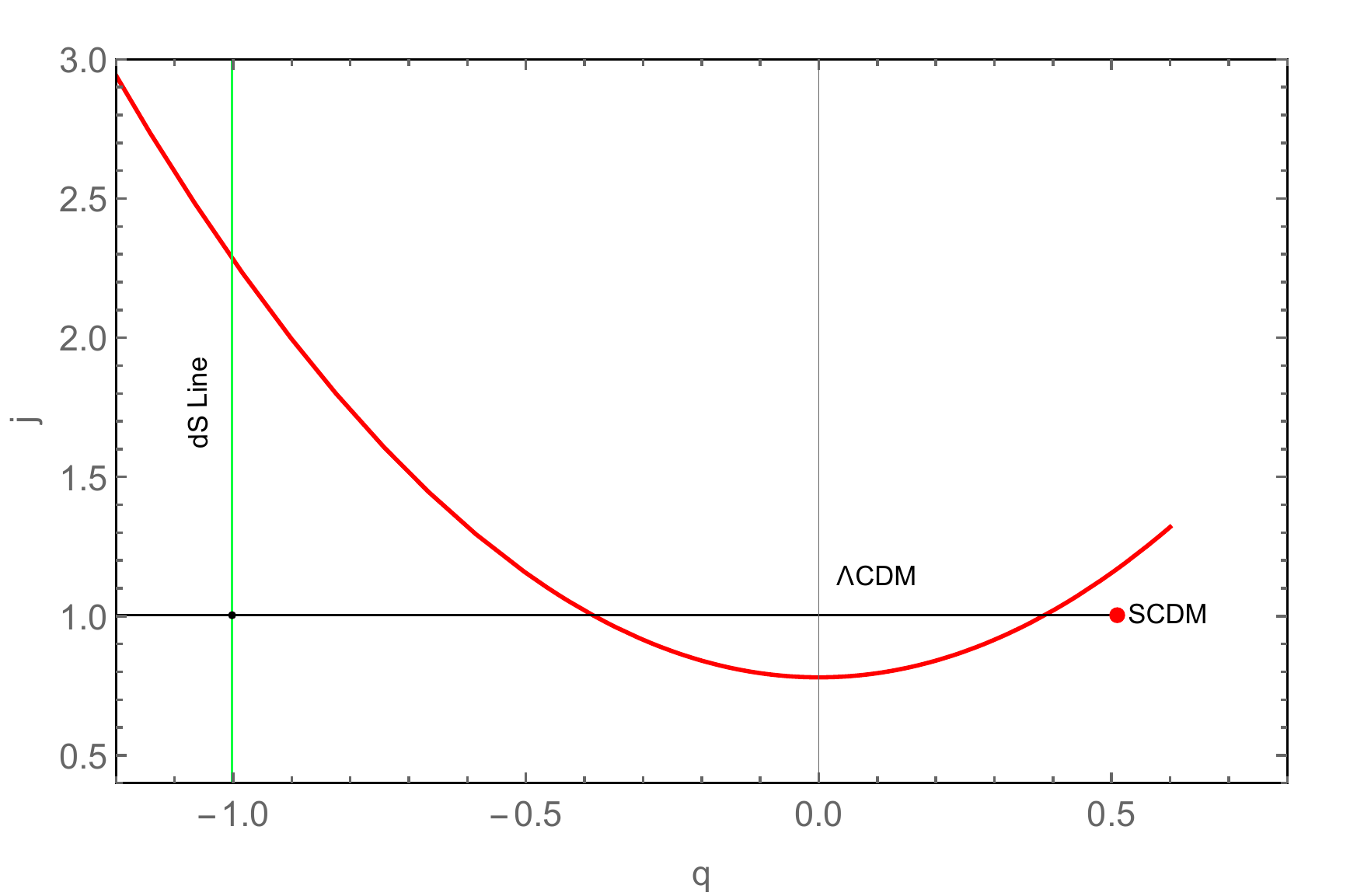}
\caption{Variation of $q$ versus $j$ with $m=1.6$ and $k=0.097$.}\label{label-11}
\end{minipage}\qquad
\begin{minipage}[b]{.4\textwidth}
\includegraphics[width=60mm]{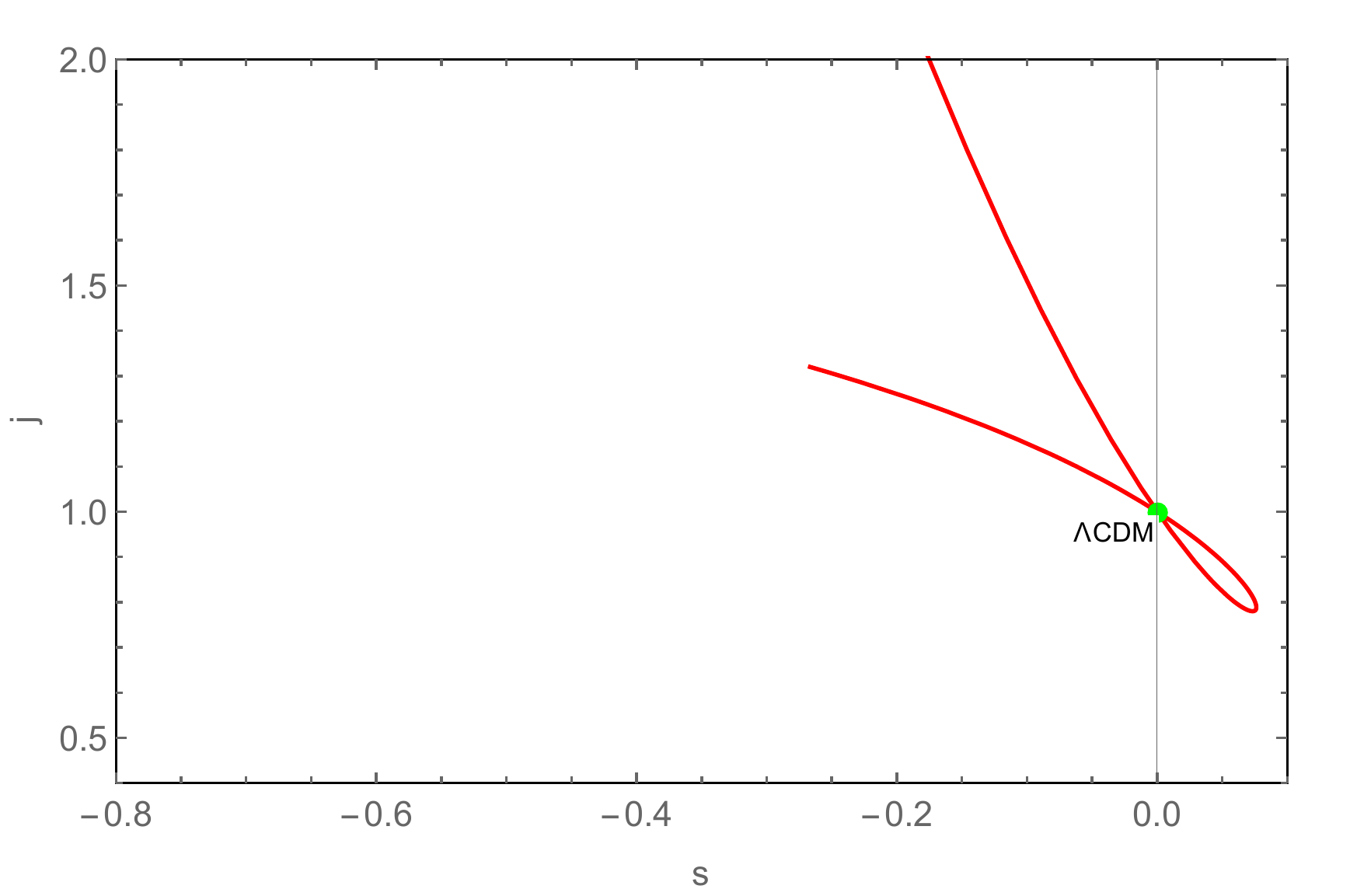}
\caption{variation of $s$ versus $j$ with $m=1.6$ and $k=0.097$.}\label{label-12}
\end{minipage}
\end{figure*}

In the Fig. 11 the vertical line is the de Sitter (dS) state at $q=-1$. The LVDP $q-j$ curve in the figure 11 crosses the dS line and going up due to Big Rip. Similarly, the LVDP $s-j$ curve crosses the $\Lambda$CDM statefinder pair $(0,1)$ two times as shown in figure 12. We observe from both the figures that our LVDP model evolve and crosses the de Sitter line and reaches to the super-exponential expansion.

\section{Conclusion}
In this study, we have investigated LRS Bianchi type I universe model with magnetized strange quark matter in $ f(R,T)$ gravitation theory for $f(R,T)=R+2f(T)$ model.

To obtain exact solutions, we have used the Hubble parameter, EoS for strange quark matter and linearly varying deceleration parameter. It is clearly seen that magnetic flux is effective and non-vanish for LRS Bianchi type I universe model and changes with cosmic time.

Also, we get $h^2\rightarrow 0$ when $t\rightarrow \infty$. This result can be interpreted as the end of the universe as the magnetic field may lose its effect. In this model, Bag constant $ B_{c} $ is effective on pressure, density and cosmological constant. While the Bag constant $B_{c}$ increases the density, it decreases the cosmic pressure value. When $t\rightarrow 0$, we get constant density also when \textit{t} increases, we obtain $\rho\rightarrow B_c$. From this we obtain small, constant and negative cosmological constant value as $\Lambda=-(8\pi+4\mu)B_c$. When \textit{t} increases we get negative pressure value, i.e., $p \rightarrow -B_c$.

On the basis of results, we can claim that strange quark matter may be source of dark energy also agree with strange quark stars because of the obtained constant pressure and density in this model. However, these results are compatible with the previous study of \cite{can} in $f(R,T)$ gravitation theory.

If we take $\mu=0$, we get $f(R,T)=R$. Then we obtain general relativity results for LRS Bianchi type I universe with magnetized strange quark matter. From eqs.(37), we obtain the magnetic flux as follows,

\begin{equation}
h^2=\frac{3(n-1)(-kt+m-3)}{(8\pi)(n+2)}\left(\frac{-2}{kt(t-t_R)}\right)^2
\end{equation}

Using eqs. (38) and (39), we get energy density and pressure for this model as follows

\begin{equation}
\rho=-\frac{3}{(8\pi)}\biggl[\frac{9(n-1)}{(n+2)^2}+\frac{3[3-3n+(-kt+m)n]}{(n+2)}\biggr]H^2+B_c
\end{equation}
\begin{equation}
p=-\frac{1}{(8\pi)}\biggl[\frac{9(n-1)}{(n+2)^2}+\frac{3[3-3n+(-kt+m)n]}{(n+2)}\biggr]H^2-B_c
\end{equation}

and from eq.(41), we find cosmological constant value in general relativity as follows

\begin{equation}
\Lambda=\biggl[\frac{3[(12n\pi+24\pi)(-kt+m-1)]}{8\pi(n+2)^2}+\frac{-19}{2(n+2)^2}\biggr]H^2-8\pi B_c
\end{equation}

 When $t\rightarrow \infty$, we get same results with f(R,T) gravitation theory. From eq.(54), we obtain magnetic flux value as $h^2\rightarrow 0$, From eq.(55), the cosmic density value as $p \rightarrow -B_c$,  from eq.(56), the cosmic density $\rho\rightarrow B_c$ and from eq.(57), we obtain different results for cosmological constant $\Lambda=-(8\pi)B_c$ in general relativity.

\end{document}